\documentclass[showpacs,showkeys,nofootinbib]{revtex4-1}

\usepackage[latin9]{inputenc}
\usepackage{amsmath}
\usepackage{amssymb}
\usepackage{graphicx}
\usepackage[unicode=true,  bookmarks=false,  breaklinks=false,pdfborder={0 0
	1},backref=section,colorlinks=false]{hyperref}
\usepackage{amsfonts}
\usepackage{subfigure}
\usepackage{cleveref}
\usepackage{listings}
\usepackage{xcolor}
\usepackage{array}
\usepackage{url}%
\hypersetup{
	colorlinks,linkcolor=blue,citecolor=blue,urlcolor=blue}
\makeatletter
\@ifundefined{textcolor}{}
{
	\definecolor{BLACK}{gray}{0}
	\definecolor{WHITE}{gray}{1}
	\definecolor{RED}{rgb}{1,0,0}
	\definecolor{GREEN}{rgb}{0,1,0}
	\definecolor{BLUE}{rgb}{0,0,1}
	\definecolor{CYAN}{cmyk}{1,0,0,0}
	\definecolor{MAGENTA}{cmyk}{0,1,0,0}
	\definecolor{YELLOW}{cmyk}{0,0,1,0}
}

\makeatother

\begin{document}
\title{Induced Gravitational Waves  from Multi-Sound Speed Resonances during Cosmological Inflation}
\author{Andrea Addazi}
\email{Addazi@scu.edu.cn}

\affiliation{Center for Theoretical Physics, College of Physics, Sichuan University,
Chengdu, 610064, PR China}
\affiliation{Laboratori Nazionali di Frascati INFN, Frascati (Rome), Italy, EU}

\author{Salvatore Capozziello}
\email{capozziello@unina.it}

\affiliation{Dipartimento di Fisica E. Pancini, Universita di Napoli Federico II and Istituto Nazionale di Fisica Nucleare, Sezione di Napoli, Compl. Univ. di Monte S. Angelo, Edificio G, Via Cinthia, I-80126, Napoli, Italy,}
\affiliation{Scuola Superiore Meridionale, Largo S. Marcellino 10, I-80138, Napoli, Italy,}

\author{Qingyu Gan}
\email{gqy@stu.scu.edu.cn}

\affiliation{Center for Theoretical Physics, College of Physics, Sichuan University,
Chengdu, 610064, PR China}
\affiliation{Scuola Superiore Meridionale, Largo S. Marcellino 10, I-80138, Napoli, Italy,}

\begin{abstract}

We explore the possibility of multi-parametric resonances from 
time varying sound speed during cosmological  inflation. 
In particular, we fix our set-up to the simpler case beyond a single oscillation model already explored in literature: 
two sinusoidal harmonics around a constant sound speed equal to one. 
 We find that,
within the perturbative regime, except for some certain extreme corners of the parameter space,
the primordial density spectrum is characterized by two groups of amplified peaks
centered around two critical oscillatory frequencies of the sound speed. 
As a general result, we show that the energy spectrum of the secondary induced GWs from the inflationary era has a single major broad peak, whereas the one from the
radiation dominated phase consists of one/two principle peak-like
configuration(s) for relatively small/large ratio of two oscillatory
frequencies. The GW relic stochastic backgrounds carry a gravitational memory 
of the parametric resonances during inflation. 
GW signals from double sound speed resonances 
can be tested in complementary channels from Pulsar-timing radio-astronomy,
space and terrestrial GW interferometers. 
\end{abstract}


\maketitle
\tableofcontents{}

\vspace{1cm}

\section{Introduction}

The detection of gravitational waves
(GWs) from mergers of binary compact
objects \cite{LIGOScientific:2016aoc,LIGOScientific:2017vwq} 
also inspire the possibility of searches for 
 stochastic GW backgrounds produced from new physics mechanisms 
in the early Universe. 
Indeed, GW physics represents a new arena for 
tests of fundamental theories in cosmology
alongside with the
cosmic microwave background (CMB) and large-scale structure (LSS).
Cosmological stochastic GW
backgrounds can be generated by 
several different sources. 
Among all the possibilities,
the secondary GW spectrum induced by primordial density perturbations in early
universe is particularly interesting and has drawn a lot of attention
recently (see for example Ref.\cite{Domenech:2021ztg} for a review in the subject). In the cosmological
perturbation theory, it is well known that the scalar and tensor perturbations
evolve independently at the linear order while they are dynamically
coupled at the second and higher orders \cite{Acquaviva:2002ud,Baumann:2009ds}.
Hence the scalar mode associated with the primordial density perturbations
can excite the tensor mode and induce the secondary GWs, when they
are either localized at scales much smaller than the Hubble horizon during the inflationary epoch
\cite{Biagetti:2013kwa,Cai:2019jah,Fumagalli:2021mpc,Peng:2021zon,Cai:2021wzd,Inomata:2021zel}
or reenter the horizon during the post-inflationary radiation dominated
(RD) and matter dominated (MD) eras \cite{Ananda:2006af,Baumann:2007zm,Assadullahi:2009nf,Kohri:2018awv}.
Such a phenomenon can generate a relic detectable stochastic GW background. 
 CMB and LSS observations imply that primordial fluctuations are distributed as approximately 
 a Gaussian profile at $\gtrsim1\textrm{Mpc}$ scale, with a near scale-invariant power spectrum
 and an amplitude that is too small to be visible in current detector.
Nevertheless,  constraints at small scales, $\ll1\textrm{Mpc}$, are less stringent and 
a large enhancement of the scalar and tensor fluctuations can be envisaged. 
Several possible mechanisms for an amplification of primordial density perturbations
and induced GWs at small scale were proposed in literature
considering local features of the inflaton potential like
near-inflection points \cite{ivanov1994inflation,Di:2017ndc,Ballesteros:2017fsr,Mahbub:2019uhl,Ragavendra:2020sop},
step-like changes \cite{Kefala:2020xsx,Inomata:2021tpx,Dalianis:2021iig}, very small periodic structure \cite{Cai:2019bmk,Peng:2021zon}
and multi-field inflation with a rapid turn \cite{Palma:2020ejf,Fumagalli:2020nvq,Fumagalli:2021mpc},
 higher dimensional operators or non-minimal coupling between inflaton and graviton
 \cite{Kannike:2017bxn,Pi:2017gih, Capozziello:2017vdi, Capozziello:2019klx, Fu:2019ttf, Ashoorioon:2019xqc, Ashoorioon:2018uey, Lin:2020goi, Kawai:2021edk},
extra axion-like curvaton or spectator scalar field \cite{Biagetti:2013kwa,Ando:2017veq,Cai:2021wzd,Inomata:2020xad,Zhou:2020kkf,Pi:2021dft,Inomata:2022ydj}, first order phase transition \cite{Khodadi:2021ees,Khodadi:2018scn, Addazi:2020zcj},
and supergravity or string inspired models \cite{Ozsoy:2020kat,Aldabergenov:2020bpt,Wu:2021zta,Zhang:2021rqs,Spanos:2021hpk,Ketov:2021fww}.

On the other hand, the sound speed of primordial perturbations
plays an important role in determining the evolution of Universe in
early epoch. Usually,  in the
standard slow-roll inflation paradigm, the sound speed of the inflaton is equal to one; but it can deviate from the
unity in alternative scenarios with non-canonical kinetic term,
e.g., k-inflation \cite{Armendariz-Picon:1999hyi,Garriga:1999vw},
Dirac-Born-Infeld inflation \cite{Silverstein:2003hf,Alishahiha:2004eh},
effective field theory from integrating out the heavy modes \cite{Achucarro:2010da,Achucarro:2010jv}
and so on. It was extensively explored in the literature that the
sizable secondary GWs at small scale can be generated by various 
behaviors of the varying sound speed of the scalar mode \cite{Wei:2004xx,Bean:2008na,Tolley:2009fg,Achucarro:2010jv,Achucarro:2010da,nakashima2011effect,Park:2012rh,Gao:2013ota,Achucarro:2014msa,Mooij:2015cxa,vandeBruck:2015tna,Cai:2015dta,Palma:2017wxu,Canas-Herrera:2020mme,Ballesteros:2021fsp, Capozziello:2018qjs}.
Recently, it was proposed that parametric resonances of 
 primordial density perturbations, originated from the
sound speed, with an oscillatory feature in of (conformal) time,
can generate a large amplification
of the abundance of induced GWs and primordial black holes (PBHs)
 in the early universe \cite{Cai:2018tuh,Cai:2019jah,Chen:2020uhe,Chen:2019zza}.
Alternatively, we mention that the parametric resonance can also be achieved
considering an oscillatory 
sound speed of the tensor perturbations during inflation \cite{Cai:2020ovp}.

In this work, we will analyze the primordial GW stochastic background 
and PBH production in case of multi-sound speed resonances. 
We will show that the GW relic signals carry 
a {\it  resonance  gravitational memory} of sound speed time-variation 
during inflation that can be tested in next generation of GW experiments. 
Multi-parametric resonances may be envisaged in many possible non-minimal coupling and multi-field inflationary dynamics in the early Universe;
thus an effective theory approach is of large interest. 
Indeed, in \cite{Cai:2018tuh}, the modification of the inflaton sound
speed by a single cosine function, with the characteristic frequency
$p_{*}$, yields resonant modes with wave-number $k$ around $p_{*}$ and a single narrow amplified peak
at $\sim p_{*}$ in the power spectrum of the primordial curvature
perturbation. 
In a more realistic scenario, it seems natural to generalize such a simple 
case to sound speed composed of many harmonics 
as a Fourier series analysis. 
First, 
in the single oscillation case, the induced GW spectrum exhibit a single principle peak-like structure and meanwhile PBHs are produced in a narrow mass spectrum. 
Indeed, secondary resonance peaks
in the source spectrum are usually sub-dominated than the major peak
by at least $\sim 5\div 6$ orders of magnitude \cite{Cai:2019jah}.
Thus, one efficient way to overcome to it is to consider a modified sound speed 
with $n$-harmonic modes with different frequencies.
In this case, one can obtain several resonance peaks 
corresponding to the $n$ frequency poles in the power spectrum. 
Accordingly, we expect to generate
the primordial density spectrum with multiple resonating peaks, enhancing the induced GWs with rich multi-peaked features and the formation
of PBHs at different mass ranges. Secondly, in  \cite{Palma:2020ejf,Fumagalli:2020adf} 
the authors showed that the primordial power spectrum can be largely
enhanced in a broad regime by a sudden turn of the trajectory in multi-field
inflation. Such a change of direction through a different slow-roll hill can cause a complicate
oscillation behavior of the sound speed \cite{Chen:2011zf,Gao:2013ota},
and it can be Fourier decomposed in a complete basis, where  each  harmonic mode might be responsible for a particular narrow resonance.

 As a first step, we focus on the double cosine parameterization,
with different amplitudes $\xi_{*1,2}$ and frequencies $p_{*1,2}$. 
In other words, here we will consider  
two dominant harmonics in the Fourier expansion, corresponding to the cases where others can be neglected.  
We will analyze the corresponding resonating pattern of the primordial curvature perturbation
and the features of the secondary GWs induced during the inflationary
and radiation dominated (RD) eras. 
We find that, within the perturbative parameter space, there are two groups of amplified peaks
around $p_{*1}$ and $p_{*2}$  in the scalar
power spectrum, which correspond to inflationary induced GWs with a single major broad peak and RD induced GWs 
with one/two principle peak-like
structures for relatively small/large ratio of $p_{*2}/p_{*1}$ (except some extreme parameter choices that will be discussed
later on in details). 
Let us mention that the double peak structure of the primordial
density power spectrum were also explored in Refs.\cite{Liu:2017hua,Liu:2018rrt,Zhang:2021vak,Wang:2021kbh,Gao:2021dfi,Zheng:2021vda}. Additionally, on top of the main features of the induced GW spectrum, we will show that there are certain fine structures produced by interactions of the multiple peaks of the source spectrum or the oscillatory
modulation of the envelope of particular resonating modes. Furthermore,  the single and double cosine oscillatory pattern of the sound speed provide us many hints for more general $n\geq3$ cosine parameterization scenarios.

The paper is organized as follows. In Sec.\ref{sec:Sound-Speed-Resonance},
we consider the Mukhanov-Sasaki equation with double oscillation of
the sound speed parameterization and we solve it numerically obtaining
the power spectrum of the comoving curvature perturbation.
Then, in Sec.\ref{subsec:R.D.-era}, we show the energy spectra of secondary GWs induced by the
primordial scalar sources during the inflationary and RD eras, for
various parameter sets respectively. Considering ongoing and forthcoming GW experiments, we study the phenomenological implications on induced GWs  from  both inflationary
and RD epochs in Sec.\ref{sec:Cosmological-Imp}. Finally, we conclude with final comments and discussions in Sec.\ref{sec:Discussion-and-conclusion}.

\section{Sound Speed Resonance and Power Spectrum}

\label{sec:Sound-Speed-Resonance}

Let us consider the Mukhanov-Sasaki equation in the Fourier space:
\begin{equation}
v_{p}^{\prime\prime}+\left(c_{s}^{2}p^{2}-\frac{z^{\prime\prime}}{z}\right)v_{p}=0,\label{eq:MS}
\end{equation}
where $v''$  second derivatives with respect to the conformal
time $\tau$ with $d\tau\equiv dt/a$. 
As a convention we assume that $\tau<0$ corresponds to 
during inflation era and $\tau>0$ is at post-inflation. 
In MS equation we introduced the  canonical variable $v\equiv z\zeta$
where $\zeta$
is the comoving curvature perturbation and $z\equiv\sqrt{2\epsilon}a/c_{s}$,
with the Hubble slow-roll parameter $\epsilon\equiv-\frac{1}{H^{2}}\frac{dH}{dt}$.
In dS approximation during inflation,  the scale factor is $a=e^{Ht}$
with constant $H$, which is equivalent to $a=-1/(H\tau)$;
$c_{s}\equiv c_{s}(\tau)$ is the time varying sound speed. 
Notice that  $\epsilon$ is assumed to be time-independent and $\epsilon<<1$ in this paper.
Thus Eq.\eqref{eq:MS} corresponds to 
\begin{equation}
v''_{p}(\tau)+\left(c_{s}^{2}p^{2}-\frac{2c'_{s}}{\tau c_{s}}-2\left(\frac{c'_{s}}{c_{s}}\right)^{2}+\frac{c''_{s}}{c_{s}}-\frac{2}{\tau^{2}}\right)v_{p}(\tau)=0.\label{eq:MS2}
\end{equation}

In Ref. \cite{Cai:2018tuh}, it was proposed that the primordial density perturbations can be overly amplified in a narrow band through a  modification with a single cosine function  to the sound speed of the inflaton. As mentioned in the introduction, in this paper we generalize it to the scenario with the sound speed of a single inflaton with multi-oscillations as follows:
\begin{equation}
c_{s}^2= 1 - 2\sum_{N}\xi_{*N}\left[1-\cos\left(2p_{*N}\tau\right)\right]\, ,\,\tau_{0}\leq\tau<0\, ,\label{eq:n-cos-cs}
\end{equation}
where $\xi_{*N}$ are small dimensionless parameters and $p_{*N}$
are the oscillatory frequencies, $N=0,1,2,...$, and $\tau_{0}$ corresponds to the conformal time when oscillation effectively starts. 
In the following, we will consider the simplified case
where the sum can be perturbatively truncated just to two harmonics
as follows:
\begin{equation}
c_{s}^{2}=1-2\xi_{*1}\left[1-\cos\left(2p_{*1}\tau\right)\right]-2\xi_{*2}\left[1-\cos\left(2p_{*2}\tau\right)\right],\,\,\tau_{0}\leq\tau<0\, .\label{eq:double-cs}
\end{equation}
Notice that Eq.\eqref{eq:double-cs} reduces to the conventional case studied in Ref.\cite{Cai:2018tuh} for vanishing $\xi_{*2}$. To have a positively definite $c_{s}$,
we take $0\leq\xi_{*1,2}\lesssim0.1$ in practice. The oscillation begins
at $\tau_{0}$, when the characteristic  $p_{*1,2}$-modes are well inside the Hubble radius,
i.e. $|\tau_{0}p_{*1,2}|>>1$. For simplicity, we assume that $c_{s}$
can transit from unity to Eq.\eqref{eq:double-cs} smoothly at the
start of the oscillation. Putting  Eq.\eqref{eq:double-cs}
into Eq.\eqref{eq:MS2} and setting the mode function at the beginning
of resonance to the Bunch-Davies vacuum $v_{p}(\tau_{0})=e^{-ip\tau_{0}}/\sqrt{2p}$,
one can numerically solve $v_{p}(\tau)$ during inflation $\tau_{0}\leq\tau<0$.

With the rapid expansion during the inflation process, the perturbative $p$-mode will eventually exit the Hubble horizon, decohering  the quantum fluctuation into the classical one. By the definition of dimensionless primordial power
spectrum, $P_{\zeta}\equiv p^{3}\left|\zeta_{p}\right|^{2}/\left(2\pi^{2}\right)$,
and the fact that it freezes at the super-horizon scale, we can rewrite
$P_{\zeta}$ in terms of $v$ as 
\begin{equation}
P_{\zeta}(p) =  2A_{s}p^{3}\left|v_{p}(\tau_{p})\tau_{p}\right|^{2},\label{eq:source-ps}
\end{equation}
where $\tau_{p}\simeq-1/(c_{s}p)\simeq-1/p$ is the sound horizon
(approximate Hubble horizon) exiting time of the corresponding $p$-mode.
Here, $A_{s}=H^{2}/(8\pi^{2}\epsilon)$ is the amplitude of
the power spectrum predicted by the conventional inflationary paradigm. From dimensional analysis, as we rescale $p\rightarrow p/\alpha,\tau\rightarrow\alpha\tau$
with constant $\alpha$, the $p$-mode scales as $v_{p}\rightarrow\sqrt{\alpha}v_{p}$
and hence $P_{\zeta}(p)$ remains the same as expected.
Thus, all physical quantities, such as $p$, $\tau$ etc., can be rescaled to  dimensional ones
 by considering $\alpha$ as a dimensional quantity. We will restore the appropriate units for each quantity in Sec. \ref{sec:Cosmological-Imp} for comparisons with experiments. 

\begin{figure}
\begin{centering}
\includegraphics[scale=0.96]{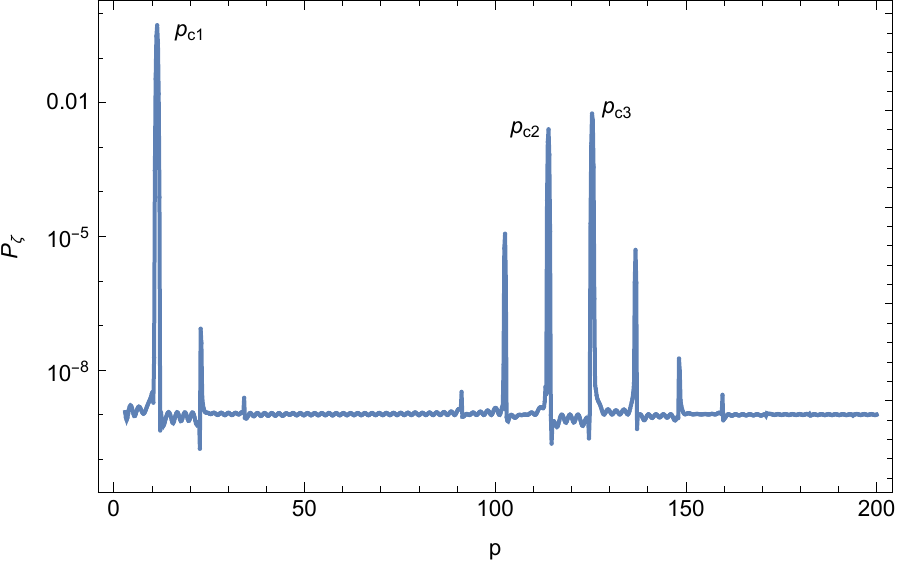}\includegraphics[scale=0.96]{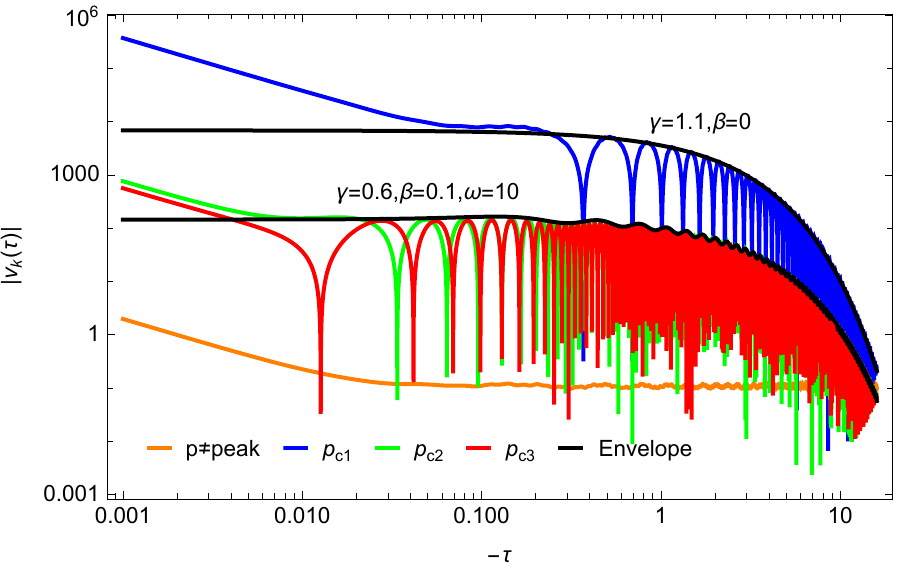}
\par\end{centering}
\caption{The power spectrum of the comoving curvature perturbation $P_{\zeta}(p)$
(Left), as a function of the peak value, and the absolute of $v_{p}(\tau)$ solutions, as a function of conformal time $\tau$, for several resonating modes (Right) with parameter set $p_{*1}=10$, $p_{*2}=100$, $\xi_{*1}=0.1$, $\xi_{*2}=0.012$, $\tau_{0}=-16$. 
In the left panel, two groups of peaks, one centering on $p\sim10$
and the other on $p\sim120$, correspond to the sound speed resonances
of characteristic oscillation modes with frequencies $p_{*1}$ and
$p_{*2}$, respectively. In the right panel, we display the $v$ solutions
corresponding to dominant peak(s) among each $p_*$-peak group, namely $p_{c1}$(blue),
$p_{c2}$(green), and $p_{c3}$(red), which are indeed boosted comparing
to the non-resonating mode with wave-number $p\protect\neq peak$
(orange). Interestingly, small oscillatory modulations are observed in the envelopes
of $|v_{p}|$ in $p_{c2}$- and $p_{c3}$-modes. The $\gamma$, $\beta$
and $\omega$ are used to parameterize the resonating modes' envelopes
given by Eq.\eqref{eq:Envelope}. }

\label{fig-vsol}
\end{figure}

In Fig.\ref{fig-vsol}, we show a typical profile of the primordial
power spectrum $P_{\zeta}(p)$, and $v_{p}(\tau)$ solutions for several resonating modes. As
a natural extend to the single cosine parameterization of $c_{s}$ investigated
in Refs.\cite{Cai:2018tuh,Cai:2019jah}, the presence of two characteristic frequencies in Eq.\eqref{eq:double-cs} results in two
groups of narrow resonating peaks centering at  $\sim p_{*1}$ and
$p_{*2}$ in the $P_{\zeta}(p)$ spectrum (see the left panel). In this sense, we simply define  ``$p_{*1}$-peak
group'' and ``$p_{*2}$-peak group'' as the 
peaks located around $\sim p_{*1}$ and $p_{*2}$ respectively. For
the case in Fig.\ref{fig-vsol}, there is only one significant peak
($p_{c1}\simeq11$) in the $p_{*1}$-peak group and two dominant peaks
($p_{c2}\simeq114$ and $p_{c3}\simeq125$) in the $p_{*2}$-peak
group, and the corresponding resonating modes are displayed in the right
panel. One can see that, the non-resonating modes with $p\neq peak$
evolve as usual in the Bunch-Davies state, whereas the resonating $p_{c1}$-, $p_{c2}$-
and $p_{c3}$-modes enter in resonance on sub-horizon scales. Besides
the similar resonance behavior as in Refs.\cite{Cai:2018tuh,Cai:2019jah} (e.g. $p_{c1}$-mode), we find new small oscillatory
modulations in the envelope of $|v_{p}|$ of $p_{c2}$- and
$p_{c3}$-modes, which are not found in the single cosine $c_{s}$ case.
The envelope of the
resonating modes before Hubble-exiting can be effectively described by the following profile:
\begin{equation}
\left|v_{p_{ci}}(\tau)\right|\propto e^{\frac{1}{2}\gamma_{i}\xi_{i}p_{ci}\left(\tau-\tau_{0}\right)}\left(1-\beta_{i}\cos\left(2\omega_{i}\left(\tau-\tau_{0}\right)\right)\right),\label{eq:Envelope}
\end{equation}
where $p_{ci}$-mode corresponds to the $i$-th peak of power spectrum.
Here, $\xi_{i}=\xi_{*1}$ ($\xi_{*2}$) if the $i$-th peak belongs
to $p_{*1}$- ($p_{*2}$-) peak group, and $\gamma_{i}$ is the correction factor
to fit the the  exponential growing magnitude of each resonating mode,
which is practically  $\sim1$ for the $p_{*1}$-peak group
and $\sim\mathcal{O}(0.1)$ for the  $p_{*2}$-peak group.
For the oscillatory modulation of the mode envelope, it can be well
matched by setting $\beta_{i}=\xi_{*1},\omega_{i}=p_{*1}$ for the $p_{*2}$-peak group and $\beta_{i}=0$
for $p_{*1}$-peak group. It is worth to
mention that such oscillatory envelope modulation of the resonating
modes is also observed in Ref.\cite{Cai:2019bmk}, where the parametric
resonance arises from a periodic structure superposing on the slow-roll inflation potential. Let us comment more on the appreance of the oscillatory behavior in the  envelope of the resonating mode in $p_{*2}$-peak group. We focus on the mode $v_{p}(\tau)$ with $p\sim p_{*2}$ and in the subhorizon regime  $ |p\tau | \gg1$. Combining Eqs. (\ref{eq:MS2})  and (\ref{eq:double-cs}) and assuming $p_{*1}\ll p_{*2}$ and $\xi_{*1}\gg \xi_{*2}$ (for most cases studied in the paper), we obtain $v_{p}^{\prime\prime}(\tau)+\left(\left(p^{2}-2p^{2}\xi_{*1}+2p^{2}\xi_{*1}\cos\left(2p_{*1}\tau\right)\right)+\left(2p^{2}\xi_{*2}-4p_{*2}^{2}\xi_{*2}\right)\textrm{cos}(2p_{*2}\tau)\right)v_{p}(\tau)=0 $. It is easy to check that in single oscillation case with $p_{*1}=p_{*2}=p_{*}$, the above equation reduces to the standard Mathieu equation $v_{p}^{\prime\prime}(\tau)+\left(\tilde{b}+\tilde{q}\textrm{cos}(2p_{*}\tau)\right)v_{p}(\tau)=0$, where $\tilde{b}$ and $\tilde{q}$ are independent of $\tau$. Whereas in the double oscillation case, the term $b\equiv p^{2}-2p^{2}\xi_{*1}+2p^{2}\xi_{*1}\cos\left(2p_{*1}\tau\right)$ depends on the variable $\tau$, so the aforementioned equation is not the Mathieu equation and can not be solved analytically. However, compared to the high-frequency oscillation term $\textrm{cos}(2p_{*2}\tau)$, the term $\cos\left(2p_{*1}\tau\right)$ oscillates much slower in frequency $p_{*1}$, indicating a slowly varying $b$ with respect to $\tau$. Such a slow oscillatory factor in Mathieu-like equation provides a clue to understand the appearance of the oscillatory modulation with the certain frequency $p_{*1}$ in the envelope of the resonating mode in $p_{*2}$-peak group.

\begin{figure}
\begin{centering}
\includegraphics[scale=0.96]{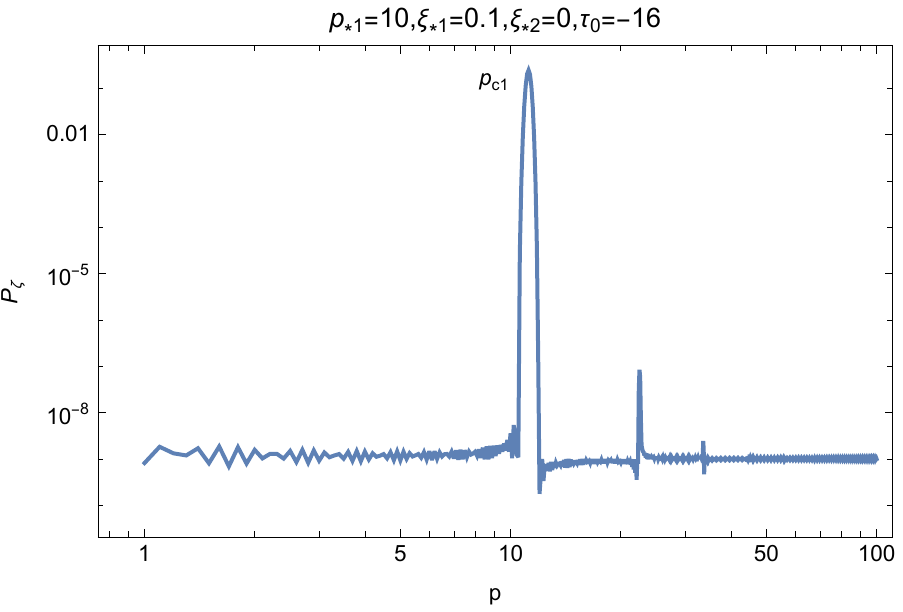}\includegraphics[scale=0.96]{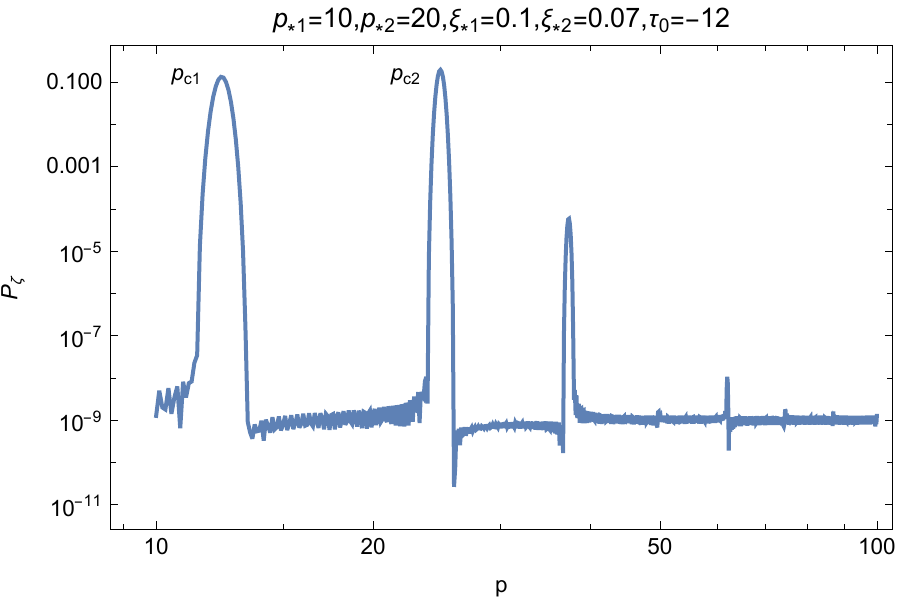}
\par\end{centering}
\begin{centering}
\includegraphics[scale=0.96]{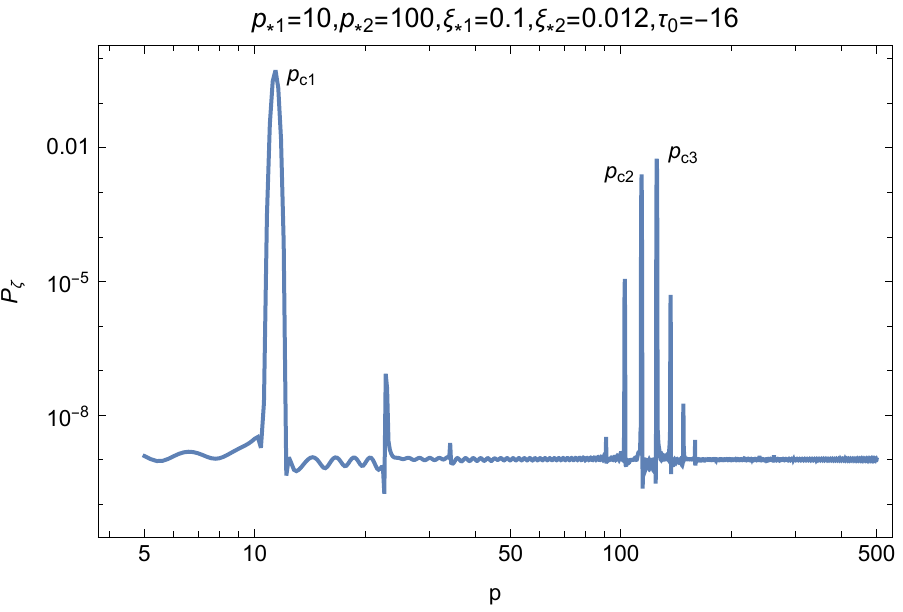}\includegraphics[scale=0.96]{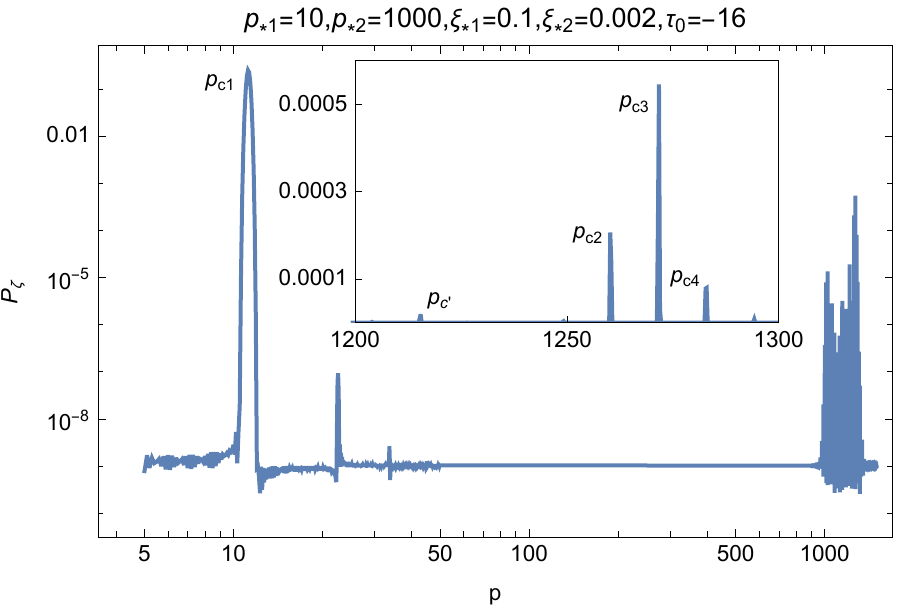}
\par\end{centering}
\caption{The power spectrum of the curvature perturbation in $(P_{\zeta},p)$
plane for different parameters. The height of all peaks in $P_{\zeta}(p)$ should be no more than
the unity in order to respect the perturbative bound. In the upper-left panel, the case
with vanishing $\xi_{*2}$ shows the conventional resonance pattern of single
cosine parameterization of $c_{s}$. In the cases with
non-vanishing $\xi_{*1}$ and $\xi_{*2}$, there can appear much more
peaks in the $p_{*2}$-peak group, especially for relatively large
ratio $p_{*2}/p_{*1}$. We define the dominant resonance peaks as those that are in the same order of amplitude of the highest peak in each $p_{*}$-peak group, and use $p_{ci}$ to denote  the $i$-th  dominant peak.
Generically, one dominant peak is found in
the $p_{*1}$-peak group and one/several dominant peak(s) in the $p_{*2}$-peak
group. Moreover, the subdominant resonance $p_{c'}$-peak in the lower-right panel gives a negligible contribution to the energy density of induced GWs, as we will see later. }

\label{fig-ps}
\end{figure}

In Fig.\ref{fig-ps}, we display the power spectra of the curvature perturbations for different parameter combinations which satisfy the perturbative condition $P_{\zeta}<1$. In particular, the $\xi_{*2}=0$
case reduces to the conventional resonance pattern predicted in Refs.\cite{Cai:2018tuh,Cai:2019jah}
with only one dominant peak $p_{c1}\sim p_{*1}$ and several  subdominant
peaks in the tail of the power spectrum. As we turn on  $\xi_{*2}$, there appears two groups of narrow resonating peaks
centering on around $p_{*1}$ and $p_{*2}$. Moreover,  much more peaks can present in the $p_{*2}$-peak group for relatively
large  ratio $p_{*2}/p_{*1}$ (see the lower panels). As we will see later, the
energy density of induced GWs is related to the scalar power spectrum
as $\Omega_{GW}\sim\int\int P_{\zeta}^{2}\sim\int\int v^{4}$ from dimensional scaling,
thus the sub-dominant peaks are more than one order
of magnitude less than the most  dominant peak in each $p_{*}$-peak group, giving
 negligible contributions to the  GW spectrum.
Hence we refer the ``dominant'' peaks to  those  in the same order of amplitude of the highest peak in each $p_{*}$-peak group, which are exclusively labeled for different cases in Fig.\ref{fig-ps}.
One can see that, except some extreme cases like vanishing $\xi_{*2}$ or small $p_{*1}\tau_0$ (see Fig.\ref{fig-Cos-imp-GW2}) ,  there is generically only one dominant peak in the
$p_{*1}$-peak group and one/several dominant peak(s) in the $p_{*2}$-peak
group.

From the resonance effect, the peaks of the power spectrum
are narrow and sharp and the power spectrum can be
approximated as sum of Dirac-delta functions
\begin{equation}
P_{\zeta}(p)  \simeq  A_{s}\left(1+\sum_{i=1}^{n}\frac{1}{2}A_{i}\xi_{i}p_{ci}\delta\left(p-p_{ci}\right)\right).\label{eq:fit-source-ps}
\end{equation}
Here, $A_{i}$ is the magnitude of the amplification
of the $i$-th resonating mode and the combination $A_{s}A_{i}$ can be
seen as the amplitude of peak's height. Actually, $A_{i}$
is reminiscent of the exponential parameterization of the power spectrum
considered in Refs.\cite{Cai:2018tuh,Cai:2019jah}. Indeed,
$\xi_{i}p_{ci}$ is determined by estimating the area of the peak
with a triangle approximation, which can be interpreted as the width
of the corresponding peak. 

\section{Induced Gravitational Waves in Early Universe}

\label{sec:Induced-Gravitational-Waves}

It is expected that the primordial density perturbations enhanced
by the narrow sound speed resonance can induce large GW signals
according to the second-order cosmological perturbation theory. In
this section, we provide theoretical analyses of the stochastic GW
background induced by the resonating modes of primordial density perturbations
during the inflationary and RD eras.

\subsection{Radiation Dominated Era}

\label{subsec:R.D.-era}

Let us compute the 
second-order tensor modes generated from the first-order
scalar perturbation with the same standard approach considered in Refs. \cite{Acquaviva:2002ud,Ananda:2006af,Baumann:2007zm}.
 In absence of anisotropy in the stress-energy
tensor, the perturbed metric in the conformal Newtonian gauge reads
\begin{equation}
d^{2}s  =  a^{2}(\tau)\left\{ -(1-2\Phi(\tau,\boldsymbol{x}))d\tau^{2}+\left[(1+2\Phi(\tau,\boldsymbol{x}))\delta_{ij}+\frac{1}{2}h_{ij}(\tau,\boldsymbol{x})\right]dx^{i}dx^{j}\right\} ,\label{eq:perturbed-metric}
\end{equation}
where $\Phi(\tau,\boldsymbol{x})$ is the Bardeen potential in first order, and
$h_{ij}(\tau,\boldsymbol{x})$ is transverse and traceless second-order tensor
mode satisfying $\delta^{ij}h_{ij}=0$ and $\delta^{ij}\partial_{i}h_{jk}=0$.
We define the Fourier transform of GW tensor $h_{ij}(\tau,\boldsymbol{x})$ as
\begin{equation}
h_{ij}(\tau,\boldsymbol{x})  \equiv  \sum_{\lambda=+,\times}\int\frac{\mathrm{d}^{3}\boldsymbol{k}}{(2\pi)^{3}}e^{i\boldsymbol{k}\cdot\boldsymbol{x}}h_{\boldsymbol{k}}^{\lambda}(\tau)e_{ij}^{\lambda}(\boldsymbol{k}),
\end{equation}
where $\lambda=+,\times$ denote the polarizations of the GWs. The
two time-independent polarization tensors $e_{ij}^{\lambda}(\boldsymbol{k})$
are expressed in terms of orthonormal basis vectors $e_{i}(\boldsymbol{k}),\bar{e}_{i}(\boldsymbol{k})$ perpendicular to $\boldsymbol{k}$,
\begin{eqnarray}
e_{ij}^{\times}(\boldsymbol{k}) & \equiv & \frac{1}{\sqrt{2}}\left[e_{i}(\boldsymbol{k})\bar{e}_{j}(\boldsymbol{k})+\bar{e}_{i}(\boldsymbol{k})e_{j}(\boldsymbol{k})\right],\nonumber \\
e_{ij}^{+}(\boldsymbol{k}) & \equiv & \frac{1}{\sqrt{2}}\left[e_{i}(\boldsymbol{k})e_{j}(\boldsymbol{k})-\bar{e}_{i}(\boldsymbol{k})\bar{e}_{j}(\boldsymbol{k})\right].
\end{eqnarray}
Sourced by the scalar perturbation $S_{lm}(\tau,\boldsymbol{x})$, the
equation of motion for the induced GWs at second order reads
\begin{equation}
h_{ij}^{\prime\prime}(\tau,\boldsymbol{x})+2\mathcal{H}h_{ij}^{\prime}(\tau,\boldsymbol{x})-\nabla^{2}h_{ij}(\tau,\boldsymbol{x})  =  -4\hat{\mathcal{T}}_{ij}^{lm}S_{lm}(\tau,\boldsymbol{x}),
\end{equation}
where the projector $\hat{\mathcal{T}}_{ij}^{lm}$ selects the transverse
and traceless parts of the source term $\hat{\mathcal{T}}_{ij}^{lm}\mathcal{S}_{lm}(\tau,\boldsymbol{x})\equiv\underset{\lambda}{\sum}\int\frac{\mathrm{d}^{3}\boldsymbol{k}}{(2\pi)^{3}}e^{i\boldsymbol{k}\cdot\boldsymbol{x}} \notag  \\ 
 e_{ij}^{\lambda}(\boldsymbol{k})e_{lm}^{\lambda}(\boldsymbol{k})\mathcal{S}_{lm}(\tau,\boldsymbol{k})$.
In the momentum space, we have
\begin{equation}
h_{\boldsymbol{k}}^{\lambda''}(\tau)+2\mathcal{H}h_{\boldsymbol{k}}^{\lambda'}(\tau)+k^{2}h_{\boldsymbol{k}}^{\lambda}(\tau)  =  S_{\boldsymbol{k}}^{\lambda}(\tau)\label{eq:GWeom}
\end{equation}
with $S_{\boldsymbol{k}}^{\lambda}(\tau)\equiv-4e_{lm}^{\lambda}(\boldsymbol{k})\mathcal{S}_{lm}(\tau,\boldsymbol{k})$.
The solution for the GWs can be obtained by using Green's function
method, 
\begin{equation}
h_{\boldsymbol{k}}^{\lambda}(\tau)  =  \frac{1}{a(\tau)}\int^{\tau}\mathrm{d}\tau_{1}a(\tau_{1})g_{\boldsymbol{k}}\left(\tau,\tau_{1}\right)S_{\boldsymbol{k}}^{\lambda}\left(\tau_{1}\right),\label{eq:h-solution}
\end{equation}
where the Green function $g_{\boldsymbol{k}}\left(\tau,\tau_{1}\right)$
in RD era is 
\begin{equation}
g_{\boldsymbol{k}}\left(\tau,\tau_{1}\right)=\frac{1}{k}\sin\left(k\tau-k\tau_{1}\right)\Theta\left(\tau-\tau_{1}\right)\label{eq:g-propagator}
\end{equation}
with  the Heaviside step
function $\Theta\left(\tau-\tau_{1}\right)$. The source term during the RD era is given
by
\begin{equation}
\mathcal{S}_{ij}(\tau,\boldsymbol{x})  =  4\Phi\partial_{i}\partial_{j}\Phi(\tau,\boldsymbol{x})+\partial_{i}\Phi\partial_{j}\Phi(\tau,\boldsymbol{x})-2\mathcal{H}^{-1}\partial_{i}\Phi\partial_{j}\Phi^{\prime}(\tau,\boldsymbol{x})-\mathcal{H}^{-2}\partial_{i}\Phi^{\prime}\partial_{j}\Phi^{\prime}(\tau,\boldsymbol{x}),\label{eq:souce}
\end{equation}
where $\mathcal{H}\equiv a'/a$ is the comving Hubble parameter related
to $H$ by $\mathcal{H}= aH$. Then from the definition of $S_{\boldsymbol{k}}^{\lambda}(\tau)$,
we obtain {\small{}
\begin{equation}
S_{\boldsymbol{k}}^{\lambda}(\tau)  =  4\int\frac{\mathrm{d}^{3}\boldsymbol{p}}{(2\pi)^{3}}e^{\lambda}(\boldsymbol{k},\boldsymbol{p})\left[3\Phi_{\boldsymbol{p}}(\tau)\Phi_{\boldsymbol{k}-\boldsymbol{p}}(\tau)+\mathcal{H}^{-2}\Phi_{\boldsymbol{p}}^{\prime}(\tau)\Phi_{\boldsymbol{k}-\boldsymbol{p}}^{\prime}(\tau)+\mathcal{H}^{-1}\Phi_{\boldsymbol{p}}^{\prime}(\tau)\Phi_{\boldsymbol{k}-\boldsymbol{p}}(\tau)+\mathcal{H}^{-1}\Phi_{\boldsymbol{p}}(\tau)\Phi_{\boldsymbol{k}-\boldsymbol{p}}^{\prime}(\tau)\right],\label{eq:effective-souce}
\end{equation}
}where $e^{\lambda}(\boldsymbol{k},\boldsymbol{p})\equiv e_{ij}^{\lambda}(\boldsymbol{k})p_{i}p_{j}$
take the explicit form $e^{+}(\boldsymbol{k},\boldsymbol{p})=p^{2}\sin^{2}\theta\cos2\varphi /\sqrt{2} $
and $e^{\times}(\boldsymbol{k},\boldsymbol{p})=p^{2}\sin^{2}\theta\sin2\varphi / \sqrt{2}$
\cite{saito2010gravitational,Espinosa:2018eve,Cai:2019jah}. Here,
$\boldsymbol{k}$ and $\boldsymbol{p}$ are respectively the wave vector of
the induced GWs and the perturbed source, and $\left(p,\theta,\varphi\right)$
are the spherical coordinates of $\boldsymbol{p}$ in system whose $\left(\hat{x},\hat{y},\hat{z}\right)$
axes are aligned with $\left(\boldsymbol{e}(\boldsymbol{k}),\bar{\boldsymbol{e}} (\boldsymbol{k}),\boldsymbol{k}\right)$.
During the RD epoch, the relation between Barden potential $\Phi_{\boldsymbol{p}}(\tau)$
and the comoving curvature perturbation $\zeta_{\boldsymbol{p}}$ is \cite{Lyth:1998xn,Ananda:2006af}
\begin{equation}
\Phi_{\boldsymbol{p}}(\tau)=\frac{2}{3}T(p\tau)\zeta_{\boldsymbol{p}},\label{eq:phi-xi}
\end{equation}
with transfer function given by
\begin{equation}
T(p\tau)=\frac{9}{\left(p\tau\right)^{2}}\left[\frac{\sin\left(p\tau/\sqrt{3}\right)}{p\tau/\sqrt{3}}-\cos\left(p\tau/\sqrt{3}\right)\right].\label{eq:transfer-func}
\end{equation}
Putting Eqs.\eqref{eq:phi-xi}-\eqref{eq:transfer-func} into
Eq.\eqref{eq:effective-souce}, one can express the source in terms
of the primordial perturbation as{\small{}
\begin{eqnarray}
S_{\boldsymbol{k}}^{\lambda}(\tau) & = & \frac{16}{9}\int\frac{\mathrm{d}^{3}\boldsymbol{p}}{(2\pi)^{3}}e^{\lambda}(\boldsymbol{k},\boldsymbol{p})f(\boldsymbol{k},\boldsymbol{p},\tau)\zeta_{\boldsymbol{p}}\zeta_{\boldsymbol{k-p}},\nonumber \\
f(\boldsymbol{k},\boldsymbol{p},\tau) & \equiv & 3T(p\tau)T(|\boldsymbol{k-p}|\tau)+\mathcal{H}^{-2}T^{'}(p\tau)T^{'}(|\boldsymbol{k-p}|\tau)+\mathcal{H}^{-1}T^{'}(p\tau)T(|\boldsymbol{k-p}|\tau)+\mathcal{H}^{-1}T(p\tau)T^{'}(|\boldsymbol{k-p}|\tau),\label{eq:souce-f}
\end{eqnarray}
}where the prime here still denotes the derivative with respect to
$\tau$.

On the other hand, the correlator for tensor metric perturbation
is defined as
\begin{equation}
\left\langle \hat{h}_{\boldsymbol{k}}^{\lambda}(\tau)\hat{h}_{\boldsymbol{k'}}^{s}(\tau)\right\rangle   =  (2\pi)^{3}\delta^{\lambda s}\delta^{(3)}\left(\boldsymbol{k}+\boldsymbol{k'}\right)\frac{2\pi^{2}}{k^{3}}P_{h}(k,\tau),\label{eq:h-correlator}
\end{equation}
where $\hat{h}_{\boldsymbol{k}}$ represents the operator associated with
the GW $\boldsymbol{k}$-mode by canonical quantization, and $P_{h}(k,\tau)$
is the dimensionless power spectrum for the GWs of each polarization.
Combining Eqs.\eqref{eq:h-solution},\eqref{eq:g-propagator},\eqref{eq:souce-f}, the GW two-point correlator can be expressed
by the primordial four-point correlator,
\begin{eqnarray}
\left\langle \hat{h}_{\boldsymbol{k}}^{\lambda}(\tau)\hat{h}_{\boldsymbol{k}^{\prime}}^{s}(\tau)\right\rangle  & = & \frac{1}{a^{2}(\tau)}\int^{\tau}\mathrm{d}\tau_{1}\int^{\tau}\mathrm{d}\tau_{2}g_{\boldsymbol{k}}\left(\tau,\tau_{1}\right)g_{\boldsymbol{k}}\left(\tau,\tau_{2}\right)a\left(\tau_{1}\right)a\left(\tau_{2}\right)\nonumber \\
 &  & \times\left(\frac{16}{9}\right)^{2}\int\frac{\mathrm{d}^{3}\boldsymbol{p}}{(2\pi)^{3}}e^{\lambda}(\boldsymbol{k},\boldsymbol{p})f(\boldsymbol{k},\boldsymbol{p},\tau_{1})\int\frac{\mathrm{d}^{3}\boldsymbol{q}}{(2\pi)^{3}}e^{s}(\boldsymbol{k'},\boldsymbol{q})f(\boldsymbol{k'},\boldsymbol{q},\tau_{2})\left\langle \hat{\zeta}_{\boldsymbol{p}}\hat{\zeta}_{\boldsymbol{k}-\boldsymbol{p}}\hat{\zeta}_{\boldsymbol{q}}\hat{\zeta}_{\boldsymbol{k}^{\prime}-\boldsymbol{q}}\right\rangle. \label{eq:h-4-corrlator}
\end{eqnarray}
Assuming the Gaussianity of the curvature perturbation, one can use
the Wick theorem and the definition of the dimensionless power
spectrum of $\zeta_{\boldsymbol{p}}$,
\begin{equation}
\left\langle \hat{\zeta}_{\boldsymbol{p}}(\tau)\hat{\zeta}_{\boldsymbol{p}^{\prime}}(\tau)\right\rangle =(2\pi)^{3}\delta^{(3)}\left(\boldsymbol{p}+\boldsymbol{p}^{\prime}\right)\frac{2\pi^{2}}{p^{3}}P_{\zeta}(p,\tau),
\end{equation}
to rewrite Eq.\eqref{eq:h-4-corrlator} as 
\begin{eqnarray}
\left\langle \hat{h}_{\boldsymbol{k}}^{\lambda}(\tau)\hat{h}_{\boldsymbol{k}^{\prime}}^{s}(\tau)\right\rangle  & = & \delta^{\lambda s}(2\pi)^{3}\delta^{(3)}\left(\boldsymbol{k}+\boldsymbol{k'}\right)\left(\frac{16}{9}\right)^{2}\frac{1}{a^{2}(\tau)}\int^{\tau}\mathrm{d}\tau_{1}\int^{\tau}\mathrm{d}\tau_{2}g_{\boldsymbol{k}}\left(\tau,\tau_{1}\right)g_{\boldsymbol{k}}\left(\tau,\tau_{2}\right)a\left(\tau_{1}\right)a\left(\tau_{2}\right)\nonumber \\
 &  & \times\int\frac{\mathrm{d}^{3}\boldsymbol{p}}{(2\pi)^{3}}e^{\lambda}(\boldsymbol{k},\boldsymbol{p})e^{s}(\boldsymbol{k},\boldsymbol{p})f(\boldsymbol{k},\boldsymbol{p},\tau_{1})f(\boldsymbol{k},\boldsymbol{p},\tau_{2})\frac{2\pi^{2}}{p^{3}}\frac{2\pi^{2}}{|\boldsymbol{k}-\boldsymbol{p}|^{3}}P_{\zeta}(p)P_{\zeta}(|\boldsymbol{k}-\boldsymbol{p}|).\label{eq:h-correlator-ps}
\end{eqnarray}
Plugging the expressions for $e^{\lambda}(\boldsymbol{k},\boldsymbol{p})$,
$g_{\boldsymbol{k}}\left(\tau,\tau_{1}\right)$ (Eq.\eqref{eq:g-propagator})
and $f(\boldsymbol{k},\boldsymbol{p},\tau)$ (Eq.\eqref{eq:souce-f}) into Eq.\eqref{eq:h-correlator-ps},
and introducing three dimensionless variables $u=|\boldsymbol{k}-\boldsymbol{p}|/k$
, $v=p/k$ and $z=k\tau$, we obtain the power spectrum of the induced
GWs after a straightforward and tedious calculation \cite{Ananda:2006af,Baumann:2007zm,Espinosa:2018eve,Bartolo:2018rku,Kohri:2018awv,Cai:2019jah,Cai:2019amo}
\begin{eqnarray}
P_{h}^{RD}(k,\tau) & = & \int_{0}^{\infty}dv\int_{|v-1|}^{(v+1)}du\left(\frac{4v^{2}-\left(1+v^{2}-u^{2}\right)^{2}}{4uv}\right)^{2}P_{\zeta}(vk)P_{\zeta}(uk)\nonumber \\
 &  & \times\frac{4}{81}\frac{1}{z^{2}}\left(\cos^{2}z\mathcal{I}_{c}^{2}(u,v,z)+\sin^{2}z\mathcal{I}_{s}^{2}(u,v,z)+\sin2z\mathcal{I}_{c}(u,v,z)\mathcal{I}_{s}(u,v,z)\right),\label{eq:ph-RD}
\end{eqnarray}
with the long expressions $\mathcal{I}_{c}$ and $\mathcal{I}_{s}$ given
by 
\begin{eqnarray}
\mathcal{I}_{c}(u,v,z) & = & 4\int^{z}\mathrm{d}\widetilde{z}(-\widetilde{z}\sin\widetilde{z})\left(2T(v\widetilde{z})T(u\widetilde{z})+\left(T(v\widetilde{z})+v\widetilde{z}T_{,1}(v\widetilde{z})\right)\left(T(u\widetilde{z})+u\widetilde{z}T_{,1}(u\widetilde{z})\right)\right),\nonumber \\
\mathcal{I}_{s}(u,v,z) & = & 4\int^{z}\mathrm{d}\widetilde{z}(\widetilde{z}\cos\widetilde{z})\left(2T(v\widetilde{z})T(u\widetilde{z})+\left(T(v\widetilde{z})+v\widetilde{z}T_{,1}(v\widetilde{z})\right)\left(T(u\widetilde{z})+u\widetilde{z}T_{,1}(u\widetilde{z})\right)\right).\label{eq:IsIc}
\end{eqnarray}
Note that $T_{,1}(x)\equiv dT(x)/dx$ for any variable $x$.

With the expansion of the universe, the induced GWs eventually  evolve into a
stochastic background which can be characterized by the energy density
fraction $\Omega_{\mathrm{GW}}(\tau,k)$, which only depends on the magnitude of the wave vector $\boldsymbol{k}$. When the relevant GW mode
is well inside the Hubble horizon during the RD era, the GW energy
density fractions related to its power spectrum as
\begin{equation}
\Omega_{\mathrm{GW}}^{RD}(\tau,k)  =  \frac{1}{24}\left(\frac{k}{\mathcal{H}}\right)^{2}\overline{P_{h}^{RD}(\tau,k)},\label{eq:GW-density-RD}
\end{equation}
where the two polarization modes have been summed over, and the overline
means the time average over several periods of the GWs \cite{Maggiore:1999vm,Boyle:2005se,Espinosa:2018eve}.
Note that $\mathcal{H}(\tau)\simeq1/\tau$ in RD epoch. To compare
with the experiments, the current abundance of GW energy spectrum
is estimated as {\small{}
\begin{eqnarray}
\Omega_{\mathrm{GW}}(\tau_{0},k)h_{0}^{2} & = & c_{g}\Omega_{\mathrm{r},0}\Omega_{\mathrm{GW}}^{RD}(\tau_{f},k)h_{0}^{2}\nonumber \\
 & = & \frac{h_{0}^{2}c_{g}\Omega_{\mathrm{r},0}}{972}\int_{0}^{\infty}dv\int_{|v-1|}^{(v+1)}du\left(\frac{4v^{2}-\left(1+v^{2}-u^{2}\right)^{2}}{4uv}\right)^{2}P_{\zeta}(vk)P_{\zeta}(uk)\left(\mathcal{I}_{c}^{2}(u,v,k\tau_{f})+\mathcal{I}_{s}^{2}(u,v,k\tau_{f})\right)\, , \label{eq:GW-density-today-RD}
\end{eqnarray}
}where the expressions of $\mathcal{I}_{c}$ and $\mathcal{I}_{s}$
are given in Eq. \eqref{eq:IsIc}. Here, $\tau_{f}$ represents the
end of the RD era when $k\tau_{f}\gg\mathcal{O}(10^{3})$ for most
of the GW $k$-mode of interest. In fact, as  discussed in
\cite{Espinosa:2018eve}, the upper and lower limits in the integration
of $\mathcal{I}_{c}$ and $\mathcal{I}_{s}$ can be safely set to
$1$ and $+\infty$, thus we use $\mathcal{I}_{s,c}(u,v,+\infty)\equiv\mathcal{I}_{s,c}(u,v)$
for simplicity in the followings. We take the present reduced dimensionless
Hubble parameter $h_{0}\simeq0.7$, radiation energy density
fraction $\Omega_{\mathrm{r},0}\simeq5.4\times 10^{-5}$ and the factor
$c_{g}\simeq0.4$ \cite{Espinosa:2018eve,Bartolo:2018rku,Pi:2020otn}.

For the source power spectrum Eq.\eqref{eq:fit-source-ps}
generated by the narrow parametric resonance  of the double cosine parameterization
of the sound speed Eq.\eqref{eq:double-cs}, we obtain the corresponding
GW energy spectrum
\begin{eqnarray}
\Omega_{\mathrm{GW}}(\tau_{0},k)h_{0}^{2} & = & \frac{h_{0}^{2}c_{g}\Omega_{\mathrm{r},0}A_{s}^{2}}{3888}\sum_{i=1}^{n}\sum_{j=1}^{n}A_{i}A_{j}\xi_{i}\xi_{j}\frac{p_{ci}}{k}\frac{p_{cj}}{k}\left(\frac{4p_{cj}^{2}k^{2}-\left(k^{2}+p_{cj}^{2}-p_{ci}^{2}\right)^{2}}{4p_{ci}p_{cj}k^{2}}\right)^{2}\nonumber \\
 &  & \times\left(\mathcal{I}_{c}^{2}(\frac{p_{ci}}{k},\frac{p_{cj}}{k})+\mathcal{I}_{s}^{2}(\frac{p_{ci}}{k},\frac{p_{cj}}{k})\right)\Theta\left(p_{ci}-|p_{cj}-k|\right)\Theta\left(p_{cj}+k-p_{ci}\right),\label{eq:GW-RD}
\end{eqnarray}
where $i,j=1,...,n$ indices correspond to the resonant peaks in source spectrum.
The  $P_{\zeta}<1$ condition implies $P_{h}<1$ or equivalently $\Omega_{\mathrm{GW}}(\tau_{0},k)h_{0}^{2}\lesssim10^{-6}$,
for the validity of the perturbative analysis.

\begin{figure}
\begin{centering}
\includegraphics[scale=0.96]{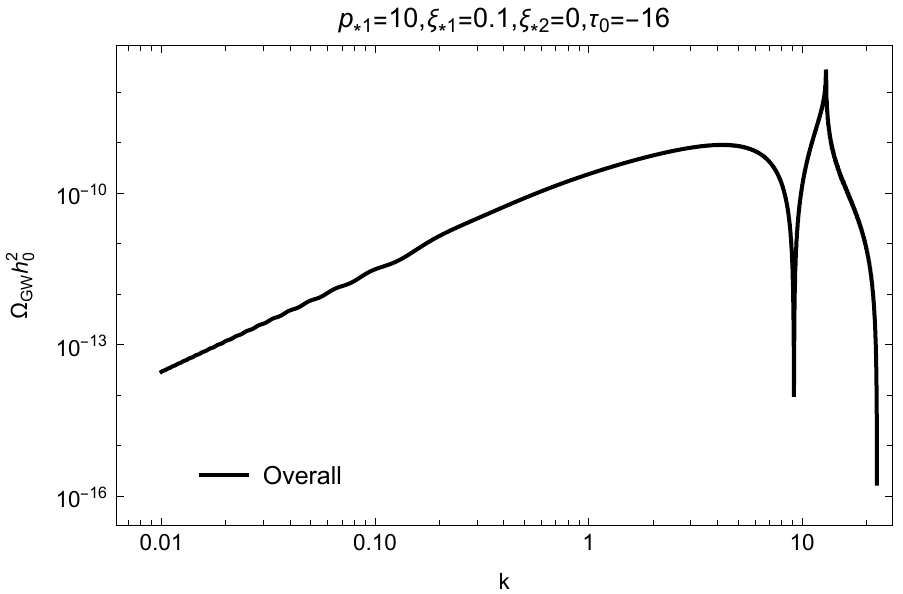}\includegraphics[scale=0.96]{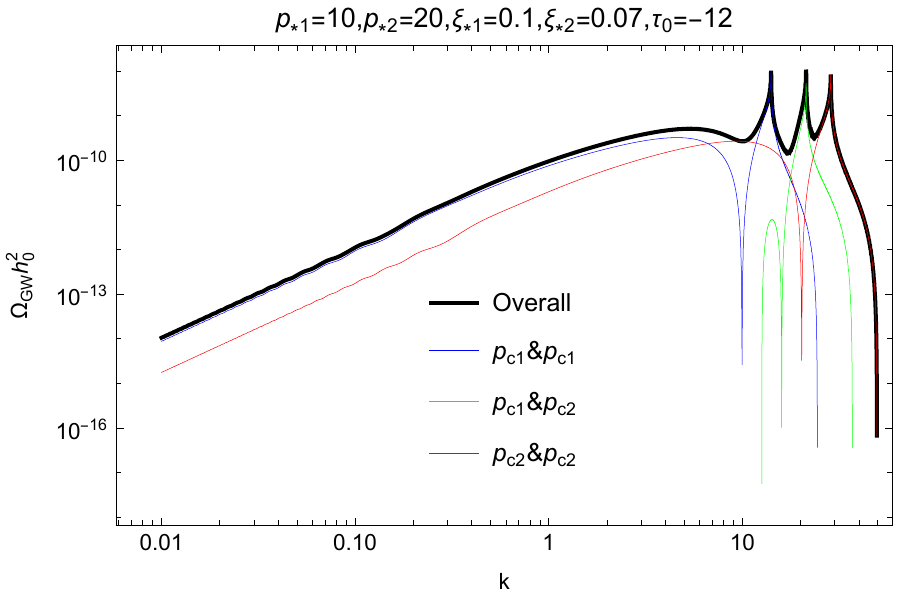}
\par\end{centering}
\begin{centering}
\includegraphics[scale=0.96]{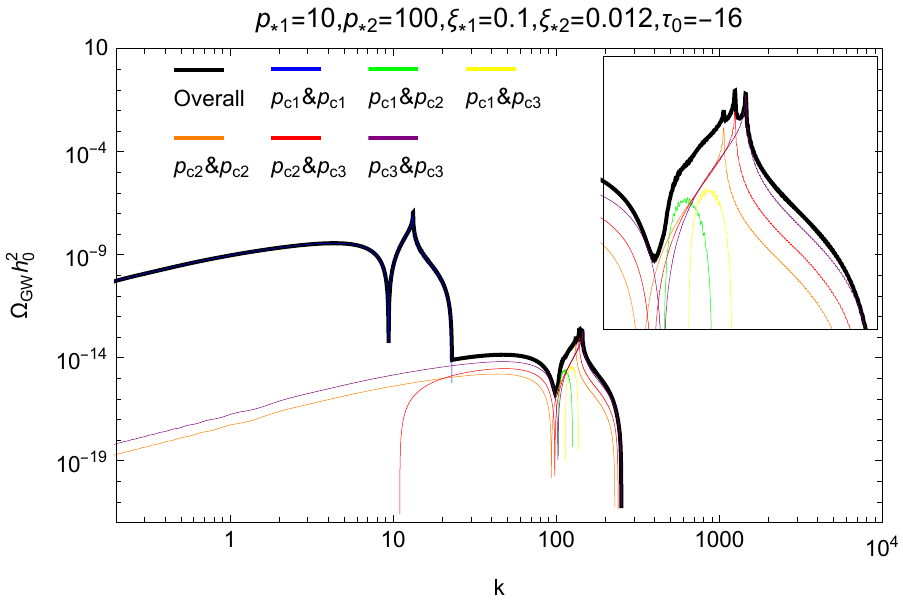}\includegraphics[scale=0.96]{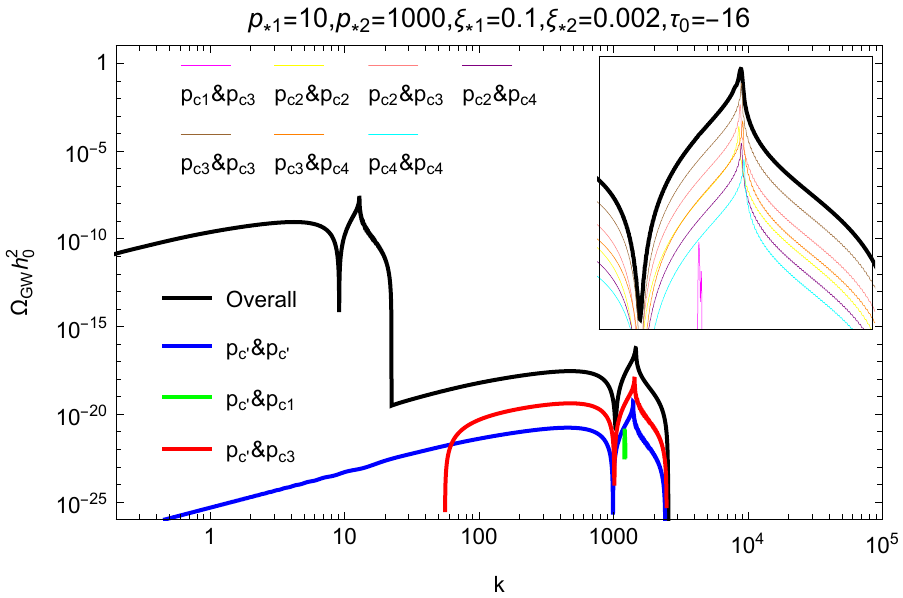}
\par\end{centering}
\caption{The energy spectrum of  GWs induced during RD epoch corresponding  to the source spectrum in Fig.\ref{fig-ps}. The black line denotes the overall
magnitude of the GW energy spectrum, and the other colored lines
denote the component decomposition generated by the convolution of
dominant  $p_{ci}$- and $p_{cj}$-peaks (see Eq.\eqref{eq:GW-RD}).
The typical profile of the GW spectrum is reproduced in
the single cosine parameterization of $c_{s}$, as displayed in
the upper-left panel. For the double cosine scenario,
the induced GW spectrum composite of at least two major peaks. Moreover, the
multiple dominant peaks in $p_{*2}$-peak group of source spectrum can give
rise to the localized fine feature with sharp spikes to the induced GW spectrum (for instance,
see inset in the lower-left panel). In the lower-right panel, one
can see that the contribution to the GW energy  from the subdominant resonating $p_{c'}$-peak (see the lower-right panel in
Fig.\ref{fig-ps}) can be safely neglected, indicating that only
dominant peaks shall be taken into account.}

\label{fig-GW-RD}
\end{figure}

In Fig.\ref{fig-GW-RD}, we display the overall induced GWs in RD
era for different parameter sets, as well as the decomposition of
each individual component corresponding to the convolution of dominant
 $p_{ci}$- and $p_{cj}$-peaks in Eq.\eqref{eq:GW-RD}. In
principle, all peaks in source spectrum $P_{\zeta}(p)$ shall be accounted, while in practice we only need to focus on the dominant
peaks, such as $p_{c1}$, $p_{c2}$ etc. as shown in Fig.\ref{fig-ps}. Indeed
 the sub-dominant peaks generate negligible contributions to the
overall energy density of the induced GWs. As shown
in the the lower-right panel, the GW spectrum components  involved with
the sub-dominant peak $p_{c'}$ (see the lower-right panel in
Fig.\ref{fig-ps}) are at least two orders of magnitude less than
the overall GW spectrum. For the case with vanishing $\xi_{*2}$
in the upper-left panel, we reproduce the conventional single-peak
GW spectrum with a sharp peak at $\simeq p_{*1}$  since
only one dominant peak is in source spectrum (see the upper-left
panel in Fig.\ref{fig-ps}). When we turn on the second oscillatory
mode with non-vanishing $\xi_{*2}$, there exist richer patterns in
GW profile governed by the amplitude and position of multiple dominant
peaks in $P_{\zeta}$. As argued in \cite{Cai:2019amo,Fumagalli:2020nvq},
the $n$  $p_{ci}$-peaks $(i=1,2...,n)$ in scalar source spectrum can generate at most $n(n+1)/2$ peaks located in
$\left(p_{ci}+p_{cj}\right)/\sqrt{3}$ in the induced GW spectrum.
In particular, two dominant peaks with nearly equal height in source
spectrum (see the upper-right panel in Fig. \ref{fig-ps}) are expected
to yield three peaks at around $p_{c1}$, $\left(p_{c1}+p_{c2}\right)/2$
and $p_{c2}$ in the GW spectrum, as shown in the upper-right panel
in Fig.\ref{fig-GW-RD}. For the cases with relatively large ratio $p_{*2}/p_{*1}$, two broad principle peak-like structures
at $\simeq p_{*1}$ and $\simeq p_{*2}$ are observed in GW spectrum, which are related to the presence of $p_{*1}$- and $p_{*2}$-peak groups
in source spectrum, respectively (see the lower panels in Figs.\ref{fig-ps} and \ref{fig-GW-RD} for a direct comparison). Since the dominant peaks at $p_{*2}$-peak
group are at $2\div3$ orders of magnitude less than the dominant
one in $p_{*1}$-peak group, the magnitude of the induced GW spectrum peak(s)
at $\simeq p_{*2}$ is less than the peak at $\simeq p_{*1}$ of
around $10$ orders. Moreover, when one zooms in the GW band at $\simeq p_{*2}$, the principle peak-like configuration can admit the localized fine structure
with several narrow peaks originated from the intersection of multiple
dominant peaks of source spectrum. Nevertheless, the appearance of
such fine structure is highly sensitive to the specific amplitudes and
positions of the resonating peaks in $P_{\zeta}$ (see the insets in
lower panels in Fig.\ref{fig-GW-RD}).

\subsection{Inflationary Era}

\label{subsec:Inflationary-era}

Now let us investigate the GWs induced by the boosted curvature
perturbations from sound speed resonances  during the inflationary
era. Due to the narrow resonance effect, GWs
can be induced by the perturbed inflaton $\delta\phi$ and its
anisotropic stress \cite{Boyle:2005se,Biagetti:2013kwa,Guzzetti:2016mkm,Cai:2019jah} as 
\begin{equation}
	\mathcal{S}_{ij}(\tau,\boldsymbol{x})=\frac{c_{s}^{2}(\tau)}{M_{p}^{2}}\partial_{i}\delta\phi(\tau,\boldsymbol{x})\partial_{j}\delta\phi(\tau,\boldsymbol{x}).
\end{equation}
In the momentum space, the effective source term in Eq.\eqref{eq:GWeom} becomes
\begin{equation}
S_{\boldsymbol{k}}^{\lambda}(\tau)  =  -4\frac{c_{s}^{2}(\tau)}{M_{p}^{2}}\int\frac{\mathrm{d}^{3}\boldsymbol{p}}{(2\pi)^{3}}e^{\lambda}(\boldsymbol{k},\boldsymbol{p})\delta\phi_{\boldsymbol{p}}(\tau)\delta\phi_{\boldsymbol{k-p}}(\tau).
\end{equation}
Performing the standard procedure as that in RD era, we first write
down the solution for the GW tensor mode in a more convenient way 
\begin{equation}
h_{\boldsymbol{k}}^{\lambda}(\tau)  =  \int^{\tau}\mathrm{d}\tau_{1}g_{\boldsymbol{k}}\left(\tau,\tau_{1}\right)S_{\boldsymbol{k}}^{\lambda}\left(\tau_{1}\right),\label{eq:h-solution-Inf}
\end{equation}
where the Green function $g_{\boldsymbol{k}}\left(\tau,\tau_{1}\right)$
during inflationary era is given by \cite{Guzzetti:2016mkm}
\begin{equation}
g_{\boldsymbol{k}}\left(\tau,\tau_{1}\right)=\frac{1}{2k^{3}\tau_{1}^{2}}e^{-ik\left(\tau+\tau_{1}\right)}\left(e^{2ik\tau}(1-ik\tau)\left(-i+k\tau_{1}\right)+e^{2ik\tau_{1}}(1+ik\tau)\left(i+k\tau_{1}\right)\right)\Theta\left(\tau-\tau_{1}\right).\label{eq:g-propagator-Inf}
\end{equation}
The GW correlator 
$\left\langle \hat{h}_{\boldsymbol{k}}^{\lambda}(\tau)\hat{h}_{\boldsymbol{k}^{\prime}}^{s}(\tau)\right\rangle$
involves the unequal-time four-point correlation function $ \left\langle \delta\hat{\phi}_{\boldsymbol{p}}(\tau_{1})\delta\hat{\phi}_{\boldsymbol{k-p}}(\tau_{1}) \right .  \delta\hat{\phi}_{\boldsymbol{q}}(\tau_{2})
\notag  \\ 
\left . \delta\hat{\phi}_{\boldsymbol{k'-q}}(\tau_{2}) \right\rangle$, 
which is  calculated in details in Ref.\cite{Figueroa:2017vfa,Caprini:2018mtu,Cai:2019cdl}.
Finally, from  Eq.\eqref{eq:h-correlator},
we obtain  the following expression for the power spectrum \cite{Biagetti:2013kwa,Fumagalli:2021mpc,Cai:2021wzd}:
\begin{equation}
P_{h}^{\textrm{Inf}}(\tau,k)  =  128A_{s}^{2}\epsilon^{2}k^{3}\int\sin^{5}\theta d\theta\int p^{6}dp\left|\int_{\tau_{i}}^{\tau}\mathrm{d}\tau_{1}g_{\boldsymbol{k}}\left(\tau,\tau_{1}\right)\tau_{1}^{2}c_{s}^{4}(\tau_{1})v_{p}(\tau_{1})v_{\boldsymbol{|k-p}|}(\tau_{1})\right|^{2},
\end{equation}
where we have used the relation $v_{p}=-\frac{1}{c_{s}M_{p}H\tau}\delta\phi_{p}$
in the spatially flat gauge. As in \cite{Cai:2019jah}, the phase
space integral $\int p^{2}\sin\theta d\theta dp$ can be integrated
out using the thin ring approximation thanks to the narrow resonance
effect. More precisely, the major contribution to $P_{h}^{\textrm{Inf}}$
comes from the quite narrow resonant regime in the neighborhood of the overly amplified
$p_{ci}$-mode (see Fig.\ref{fig-ps});
so it is reasonable to set $v_{p}=v_{p_{ci}}$ for $p\in\left(p-\frac{1}{2}\xi_{i}p_{ci},p+\frac{1}{2}\xi_{i}p_{ci}\right)$
and $v_{p}=0$ otherwise. Then the geometric interpretation of the
available phase space integral is the volume of the ringlike intersection
of two spheres: $i(j)$-sphere with radius $p_{ci}\left(p_{cj}\right)$
and thickness $\xi_{i}p_{ci}\left(\xi_{j}p_{cj}\right)$, and the
distance of centers of $i$-sphere and $j$-sphere is $k$ satisfying
$|p_{ci}-p_{cj}|<k<p_{ci}+p_{cj}$. We adopt a simple approximation
of the volume of such intersect configuration as $\Delta\Pi_{ij}\simeq2\pi p_{ci}\sin\theta_{ij}\xi_{i}p_{ci}\xi_{j}p_{cj}$
with $\theta_{ij}$ given by $\cos\theta_{ij}=(p_{ci}^{2}+k^{2}-p_{cj}^{2})/(2p_{ci}k)$.
Taking  all dominant peaks of  $P_{\zeta}$ into account, we obtain
the induced GW power spectrum at the end of inflation $\tau\simeq0$,
\begin{eqnarray}
P_{h}^{\textrm{Inf}}(\tau\simeq0,k) & \simeq & A_{s}^{2}\epsilon^{2}\frac{16}{\pi k^{3}}\sum_{i=1}^{n}\sum_{j=1}^{n}\Delta\Pi_{ij}\left(1-\left(\frac{p_{ci}^{2}+k^{2}-p_{cj}^{2}}{2p_{ci}k}\right)^{2}\right)^{2}p_{ci}^{4}\Theta(p_{ci}+p_{cj}-k)\Theta(k-|p_{ci}-p_{cj}|)\nonumber \\
 &  & \times\left|\int_{\tau_{0}}^{0}\mathrm{d}\tau_{1}e^{-ik\tau_{1}}\left(\left(-i+k\tau_{1}\right)+e^{2ik\tau_{1}}\left(i+k\tau_{1}\right)\right)c_{s}^{4}(\tau_{1})v_{p_{ci}}(\tau_{1})v_{p_{cj}}(\tau_{1})\right|^{2},\label{eq:ph-inf}
\end{eqnarray}
where the integral over $\tau_{1}$ is performed
in a numerical way. For the induced GWs generated in the inflationary
epoch, the abundance of the relic GW energy spectrum at present can
be approximated as \cite{Zhao:2006mm}
\begin{equation}
\Omega_{\mathrm{GW}}(\tau_{0},k)h^{2}  \simeq  10^{-6}P_{h}^{\textrm{Inf}}(k,\tau\simeq0).\label{eq:es-inf}
\end{equation}
Again, $P_{h}^{\textrm{Inf}}(k,\tau\simeq0)$ has to be smaller
than unity, namely $\Omega_{\mathrm{GW}}(\tau_{0},k)h^{2}\lesssim10^{-6}$,
to insure that all analysis is performed within perturbative regime.
It is worth noting that in both Eqs.\eqref{eq:GW-RD},\eqref{eq:ph-inf},
one can see that the convolution of resonating $p_{ci}$- and $p_{cj}$-
modes ($p_{ci}$ can equal to $p_{cj}$) contributes to the overall
GWs with wave-number $k$ subjected to the window $|p_{ci}-p_{cj}|<k<p_{ci}+p_{cj}$.
Moreover, just as the scale invariance of the scalar source spectrum, the GW power
spectrum $P_h$ as well as the energy density fraction $\Omega_{\mathrm{GW}}$ for both RD and
inflationary cases, i.e. Eqs.\eqref{eq:ph-RD}, \eqref{eq:GW-density-today-RD},
\eqref{eq:ph-inf} and \eqref{eq:es-inf}, all remain invariant with the
rescaling $p_{ci}\rightarrow p_{ci}/\alpha,k\rightarrow k/\alpha,\tau\rightarrow\alpha\tau$
and $v_{p}\rightarrow\sqrt{\alpha}v_{p}$.

\begin{figure}
\begin{centering}
\includegraphics[scale=0.9]{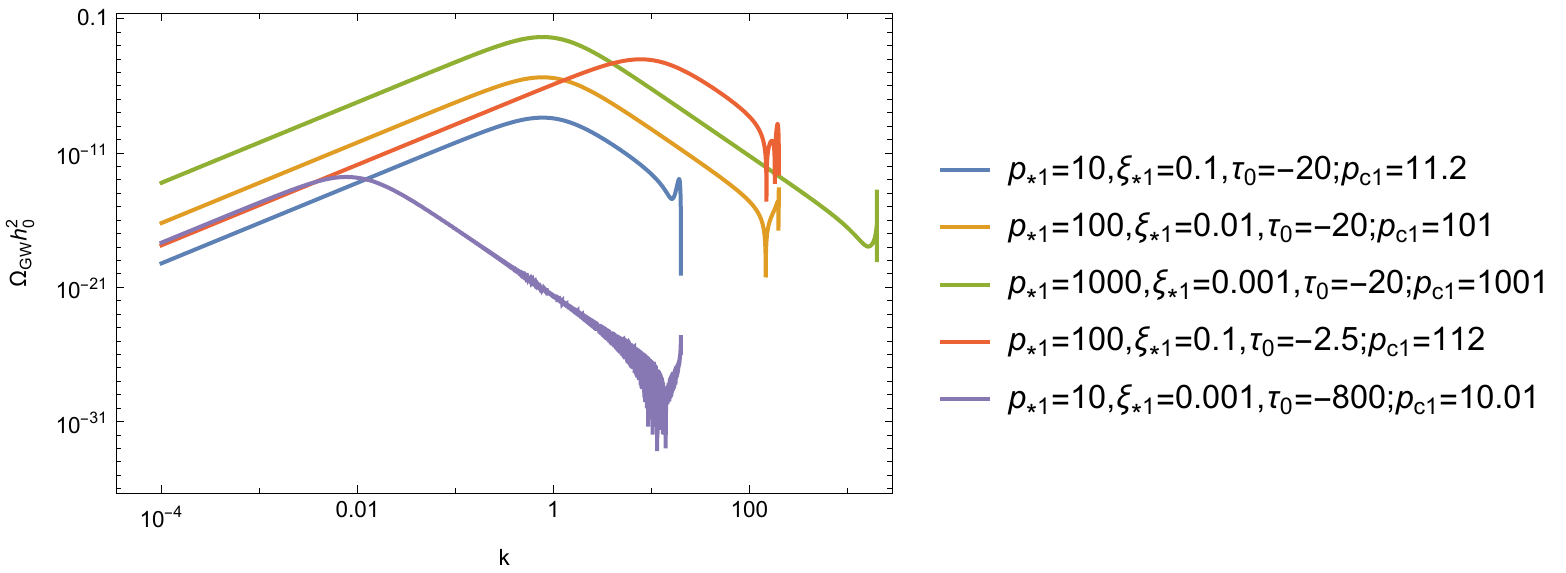}
\par\end{centering}
\caption{The energy spectrum of induced GWs from inflationary era for single
cosine $c_{s}$ parameterization  by setting $\xi_{*2}=0$. Note that here we forget
about the constraint $P_{h}^{\textrm{Inf}}<1$, thus we are free to
adjust the parameters. Similar to the upper-left panel in Fig.\ref{fig-ps},
the source spectrum $P_{\zeta}$ for each parameter set here contains
only one dominant narrow peak $p_{c1}$, which is in the neighborhood
of the characteristic scale $p_{*1}$. The spectrum curves show that
the major broad peaks of GWs induced by  certain $p_{c1}$-modes are located at $\sim p_{c1}\xi_{1}$.}

\label{fig-GW-Inf-main-peak}
\end{figure}

\begin{figure}
\begin{centering}
\includegraphics[scale=0.96]{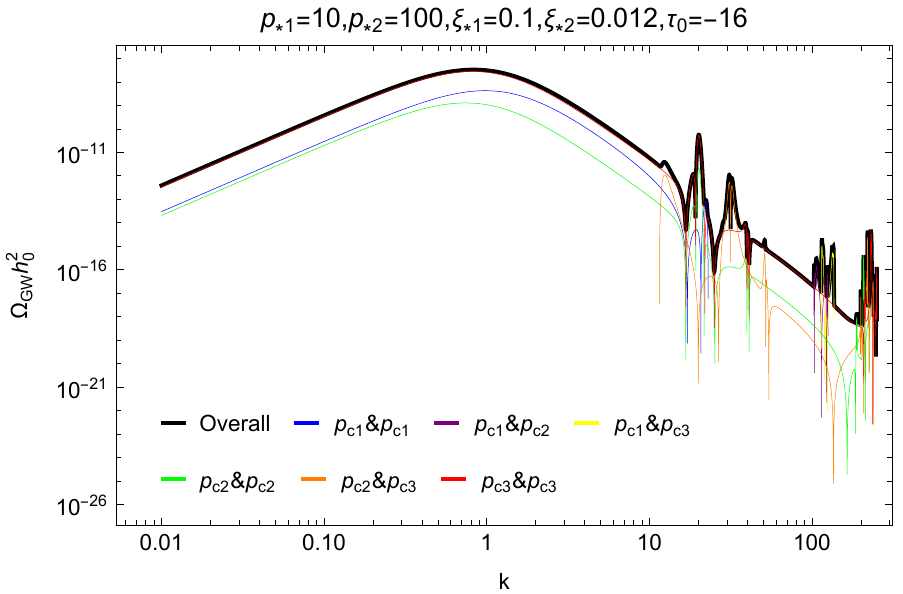}\includegraphics[scale=0.96]{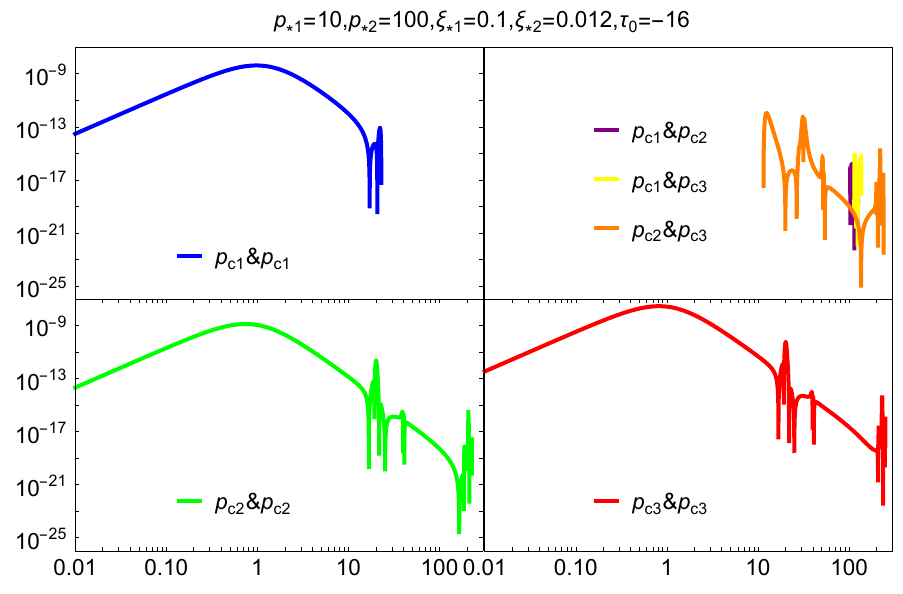}
\par\end{centering}
\begin{centering}
\includegraphics[scale=0.96]{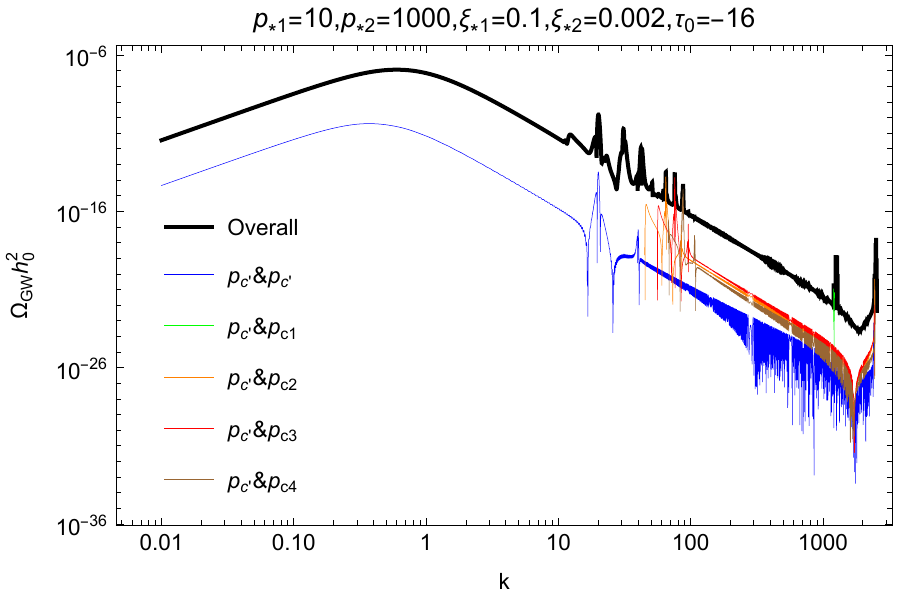}\includegraphics[scale=0.96]{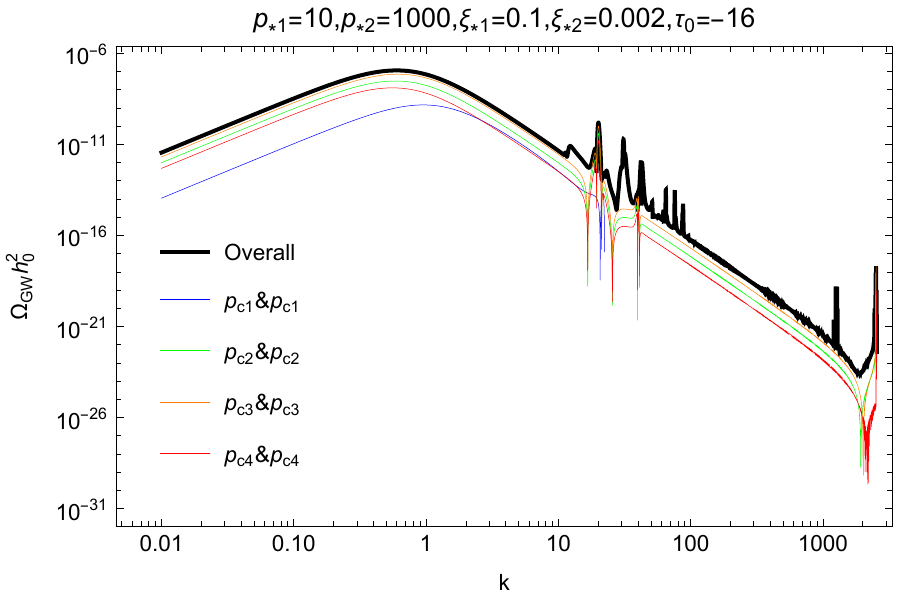}
\par\end{centering}
\caption{The energy spectrum of induced GWs from inflationary
era for  parameter sets as those in lower panels in Fig.\ref{fig-ps}. The black line denotes the overall 
GW energy spectrum, which consists of the components generated by
the convolution of dominant resonating $p_{ci}$- and $p_{cj}$-modes
(see Eq.\eqref{eq:ph-inf}). Note that each $p_{ci}$-mode corresponds
to the narrow $p_{ci}$-peak in Fig.\ref{fig-ps}. As displayed in
the lower-left panel, the components involved with the sub-dominant
$p_{c'}$-mode (see the lower-right panel in Fig.\ref{fig-ps}) has unrelevant 
contributions to the overall GW spectrum, indicating that the main features of the GW spectrum are related to
dominant $p_{ci}$-peaks. Contrary to the conventional two principle peak-like
structures at $\sim p_{*1}$ and $\sim p_{*2}$ in the GW spectrum
during RD era (see Fig.\ref{fig-GW-RD}), in the inflationary GW
spectrum there is only a single
main broad peak at $\sim p_{*1}\xi_{*1}\sim p_{*2}\xi_{*2}\sim 1$   since all $p_{ci}\&p_{ci}$-components have the major
peaks at $\sim p_{ci}\xi_{i}\sim1$ for the given parameters.
Moreover, it is interesting to observe that all $p_{ci}\&p_{ci}$-curves
with $p_{ci}$-mode belonging to the $p_{*2}$-peak group share
a common feature with a sharp spike at $\sim 2p_{*1}$, dominantly contributing
 to the first significant sharp peak in the intermediate band of the overall GW spectrum.}

\label{fig-GW-Inf}
\end{figure}

\begin{figure}
\begin{centering}
\includegraphics[scale=0.96]{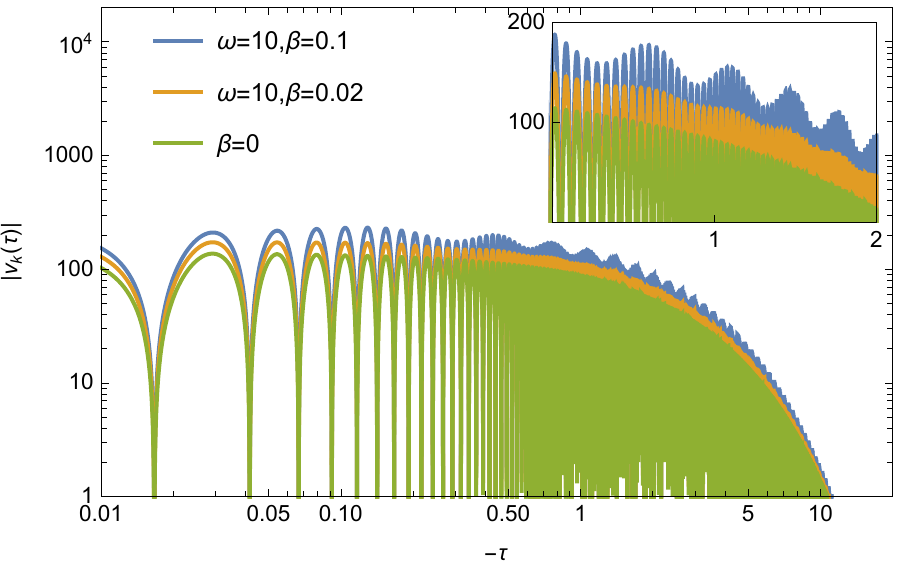}\includegraphics[scale=0.96]{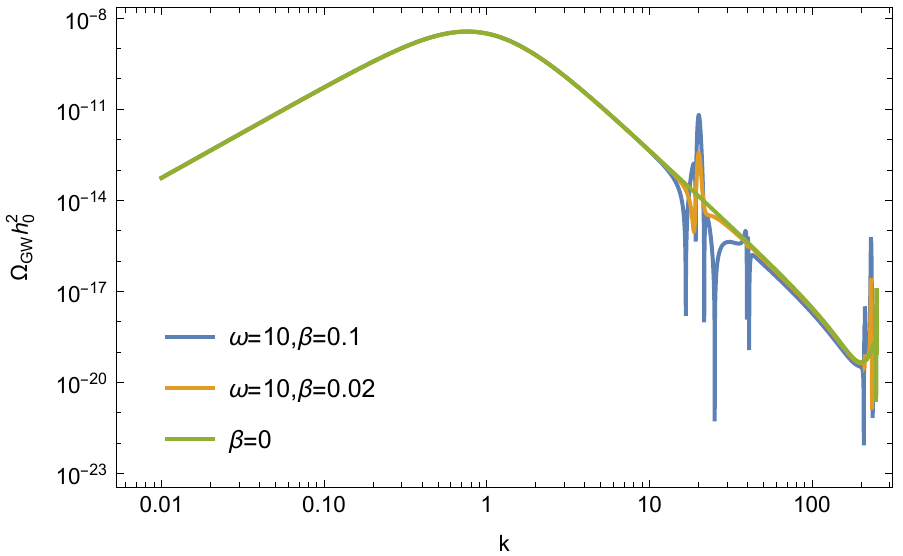}
\par\end{centering}
\begin{centering}
\includegraphics[scale=0.96]{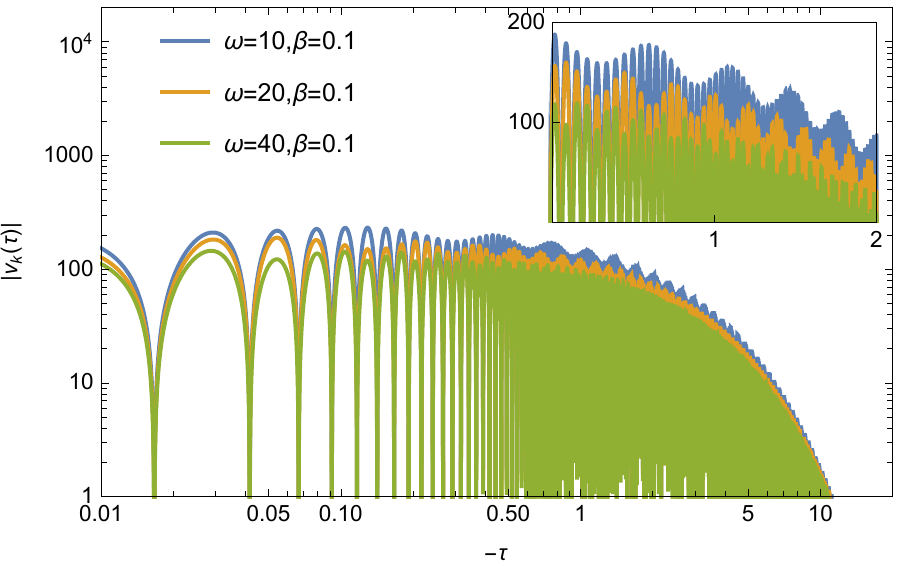}\includegraphics[scale=0.96]{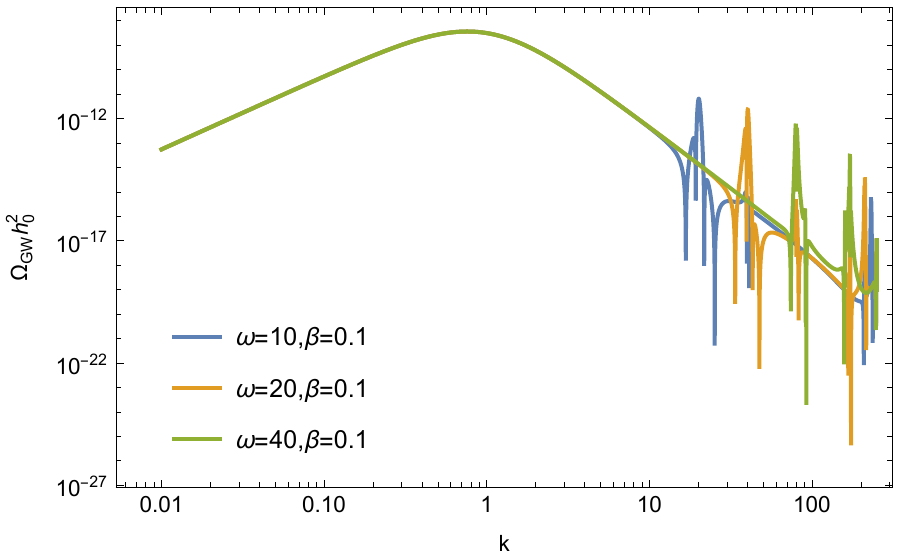}
\par\end{centering}
\caption{The resonating modes with different magnitudes and frequencies in their
envelope sinusoidal modulation (left column) and the corresponding
induced GW spectra (right column). In the left column, the blue
line represents a $|v_{p}|$ solution of the $p_{c3}$-mode shown
in Fig.\ref{fig-vsol}, whereas the other curves denote numerical solutions from Eq.\eqref{eq:Envelope}, which are set as the control group with  adjusted  shapes of
the envelopes. In the upper row,
we find that amplitude of the first sharp spike in GW spectrum is
proportional to the magnitude of the envelope oscillation, which disappears for vanishing $\beta$ parameter. Moreover, as
demonstrated in the lower row, the frequency $\omega$ of the envelope
oscillation shifts the position of  the first sharp spike in GW
spectrum as $\sim2\omega$.}

\label{fig-En-Os}
\end{figure}

We first apply Eqs.\eqref{eq:ph-inf},\eqref{eq:es-inf} to the
case with single cosine  parameterization of the sound speed,
and display the induced GW energy spectrum for different parameters
in Fig.\ref{fig-GW-Inf-main-peak}. For vanishing $\xi_{*2}$, only one narrow
dominant  $p_{c1}$-peak appears in source spectrum, and the corresponding resonating $p_{c1}$-mode
leads to  GW spectrum with a major peak at $\sim p_{c1}\xi_{1}$. In fact, this conclusion
can be generalized to the case with multiple dominant peaks appearing in double cosine
$c_{s}$ scenario;  the major broad peak of the each component of the GW spectrum
 induced by the corresponding resonating $p_{ci}$-mode is located at $\sim p_{ci}\xi_{i}$.
This is evident in Fig.\ref{fig-GW-Inf}, where we plot
overall GW spectrum consisting of some components with the convolution
of certain $p_{ci}\& p_{cj}$- modes (see Eq.\eqref{eq:ph-inf})
in inflationary era for different parameter sets.  Accordingly,
for the same parameters as  Fig.\ref{fig-ps}, the positions
of major peaks of all $p_{ci}\&p_{ci}$  components are at $\sim p_{ci}\xi_{i}\sim p_{*1}\xi_{*1}\sim p_{*2}\xi_{*2}\sim1$,
leading to a single main peak in the overall GW spectrum. This marks
a visible difference to the overall GW spectrum in RD era usually with two
principle peak-like structure at $\sim p_{*1}$ and $\sim p_{*2}$ (see
Fig.\ref{fig-GW-RD}). As mentioned in the RD case, here we also
only need to take into account the dominant peaks, since the sub-dominant 
peaks contribute little to the overall GW energy density as illustrated
in the lower-left panel in Fig.\ref{fig-GW-Inf}.  Next, let us focus  on the first sharp significant spike on the right tail of the overall inflationary GW spectrum shown in Fig. \ref{fig-GW-Inf}. One can see that such spike receives the contributions only from the $p_{ci}\&p_{ci}(i=2,3,...)$-curves, not the $p_{c1}\&p_{c1}$- and $p_{ci}\&p_{cj}(i\neq j,i,j=1,2,3,...)$-curves \footnote{Let us note that there is only one dominant resonating peak in the $p_{*1}$-peak group and normally more than one dominant resonating peak in $p_{*2}$-peak group (see Fig. \ref{fig-ps}). In order to clarify such a statement, here we denote $p_{c1}$ the dominant resonating peak in the $p_{*1}$-peak group, and $p_{ci}(i=2,3...)$ dominant resonating peaks in the $p_{*2}$-peak group. \label{footnote1}}. Let us note that $p_{c1}\&p_{c1}$-curves in the upper-left and lower-right panels do not have prominent spike at the same location of first sharp spike of the overall GW spectrum. Furthermore, one can see that the first sharp spikes of all $p_{ci}\&p_{ci}(i=2,3,...)$-curves share a common position at around $k\simeq20\simeq 2p_{*1}$. On the other hand, as displayed in Fig. \ref{fig-vsol}, one remarkable difference among the $p_{c1}$-mode and $p_{ci}(i=2,3,..)$-modes is the presence of the small oscillations in the envelope of $p_{ci}(i=2,3,..)$-mode functions, whose oscillatory frequencies $\omega_{i}$ share a common characteristic frequency $\omega_{i}=p_{*1}$ (see Eq. (\ref{eq:Envelope})). Therefore, it is quite reasonable to relate these two common phenomena, and we interpret it as the fact that the oscillatory modulation of  mode envelope generates a new resonance effect, related to the first sharp spike on inflationary GW spectrum.
In order to investigate the role of the presence of the sinusoidal modulation of the envelope, in Fig.\ref{fig-En-Os} a comparison of the GW spectra, induced by the resonating modes with different magnitudes and frequencies in their envelope oscillations, is displayed. Indeed, one can see that the amplitude of the first narrow sharp spike in GW spectrum is proportional to the magnitude of envelope's oscillation; meanwhile the  frequency $\omega$ of the envelope oscillation primarily determines the position of the sharp peak located at $\sim2\omega$.
This is exactly our interpretation fo the phenomena as in Fig. \ref{fig-GW-Inf}:  $p_{c1}\&p_{c1}$-curve does not have such spike because $p_{c1}$-mode function does not have the envelope oscillation; the first sharp spikes of all $p_{ci}\&p_{ci}(i=2,3,...)$-curves share a common position at around $k\simeq2\omega_{i}=2p_{*1}=20$. The secondary peak in GW spectrum induced by the oscillatory modulation of the envelope of the resonating modes in $p_{*2}$-peak group can be viewed as a new kind of the resonance mechanism, which is unique from the double cosine parameterization of sound speed. Moreover, as in the RD case, the intersection effect  of different $p_{ci}$- and $p_{cj}$- modes can produce multiple sharp spikes in the band  range $|p_{ci}-p_{cj}|<k<p_{ci}+p_{cj}$ of the overall GWs spectrum (see the upper-right panel in Fig.\ref{fig-GW-Inf} for instance). On the other hand, it is worth to mention that irregular rapid oscillation for the GW spectrum with much lower magnitude $\Omega_{GW}h^2_0\lesssim 10^{-25}$ is interpreted as  numerical noise, for instance, the blue  line at $k\gtrsim 200$ in the lower-left panel in Fig. \ref{fig-GW-Inf}, and the red line at $f \gtrsim 1 \textrm{Hz}$ in the lower-right panel in Fig. \ref{fig-Cos-imp-GW1}. In short, the typical profile of the inflationary GWs induced by the double sound speed resonances consists of a single broad peak at $k\sim p_{*1}\xi_{*1}\sim p_{*2}\xi_{*2}\sim 1$, followed by several significant multiple peaks on the right tail at $k\sim2p_{*1}$, and possible irregular rapid oscillation in regime with much low magnitude $\Omega_{GW}h^2_0\lesssim 10^{-25}$. 

\section{Phenomenological Implications}

\label{sec:Cosmological-Imp}

\begin{figure}
\begin{centering}
\includegraphics[scale=0.96]{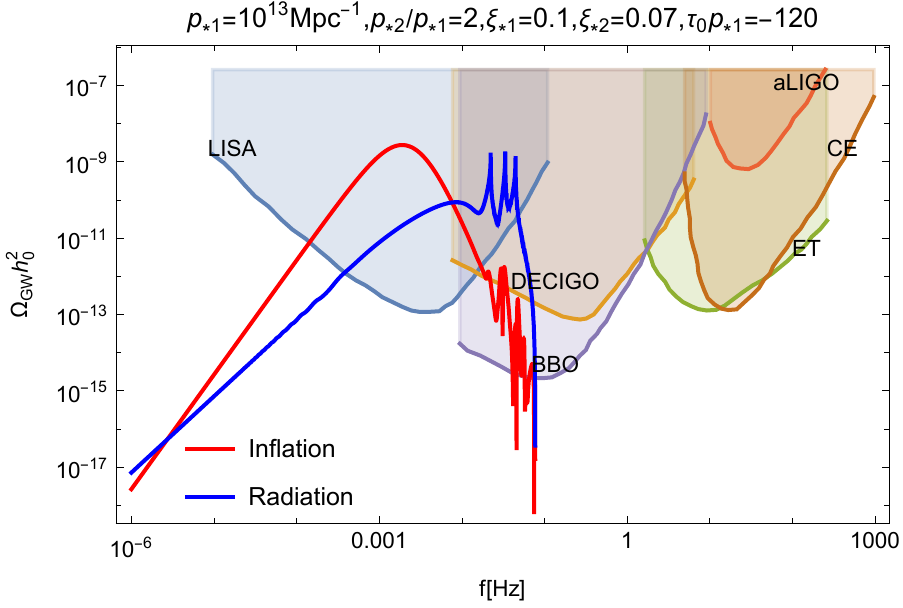}
\includegraphics[scale=0.96]{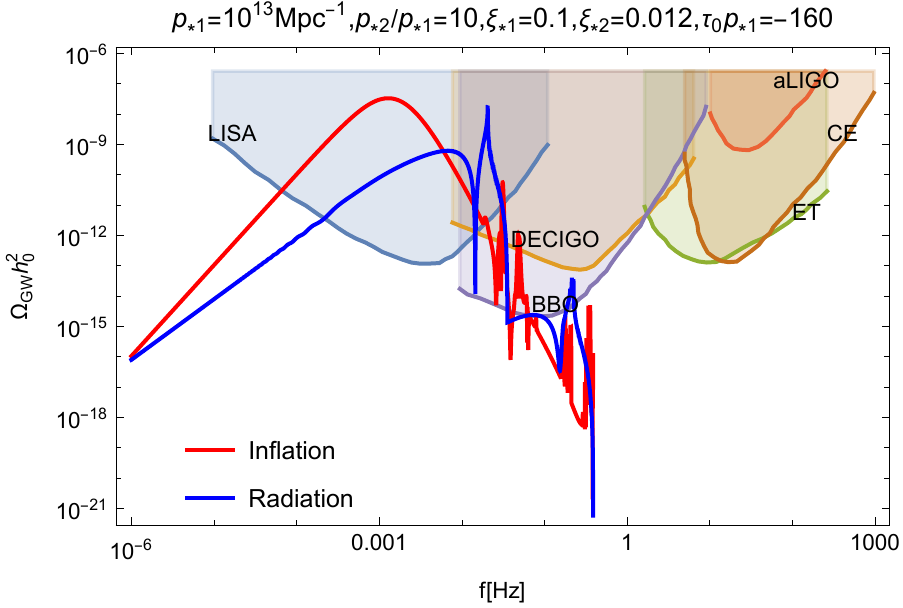}
\par\end{centering}
\begin{centering}
\includegraphics[scale=0.96]{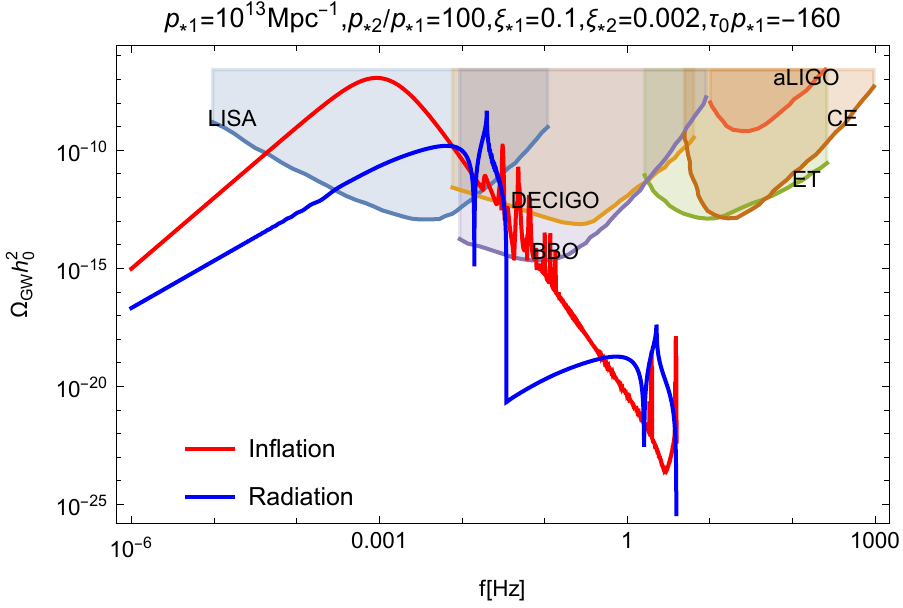}\includegraphics[scale=0.96]{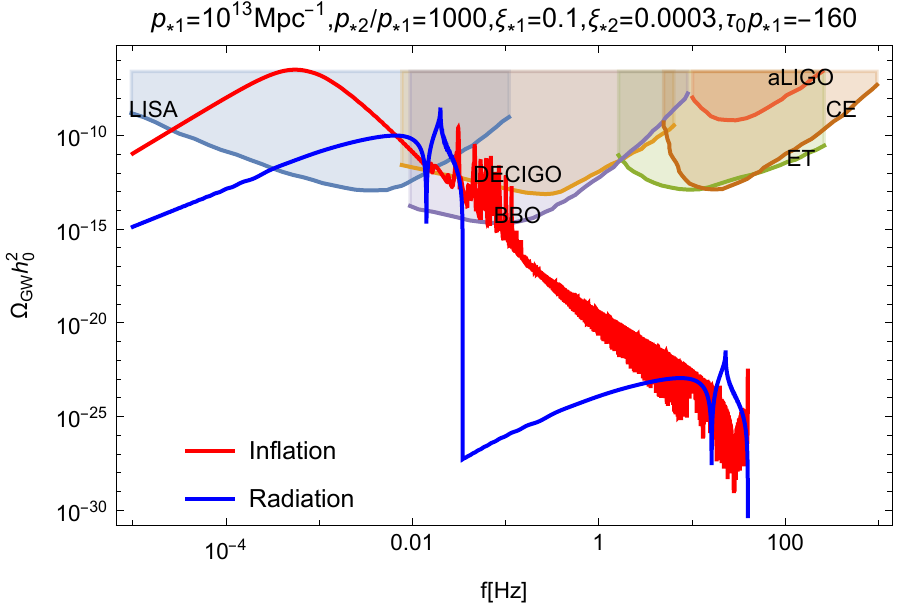}
\par\end{centering}
\caption{A comparison between GW energy spectrum with sensitivity
curves of LISA, DECIGO,
BBO, aLIGO, ET and CE. By virtue of the scaling property, a fiducial
scale is set to $p_{*1}=10^{13}\textrm{Mpc}{}^{-1}$ (i.e. $f_{*1}^{S}\simeq10^{-2}\textrm{Hz}$)
corresponding to the optimal sensitivity window of LISA, and we plot
the induced GW spectra from inflationary (red line) and RD eras (blue
line) for different ratios $r\equiv f_{*2}^{S}/f_{*1}^{S}=p_{*2}/p_{*1}$.
Notice that the parameters in $r=2,10,100$ cases are simply the rescaling
of those in Fig.\ref{fig-ps}, which indicates that the profile
of the GW spectrum shares the same features as the corresponding one
in Fig.\ref{fig-GW-RD} and Fig.\ref{fig-GW-Inf}. The abundance of the relic GWs  from inflationary era are comparable to that from the RD era. In all cases, the major peak of the inflationary
GWs spectrum located at $f^{GW}\sim f_{*1}^{S}\xi_{*1}\sim f_{*2}^{S}\xi_{*2}\sim10^{-3}\textrm{Hz}$
overlaps with the sensitivity regime of LISA, and the multiple spikes
in the intermediate band $\left(10^{-2}\textrm{Hz},10^{-1}\textrm{Hz}\right)$,
originated from the effect of envelope oscillatory modulation 
 and the intersection of multiple resonating source peaks, are in
the sensitivity window of DECIGO/BBO. For the RD GW spectrum, a
single principal peak-like structure at $f^{GW}\sim f_{*1}^{S}$
for $r=2$, whereas two principle peak-like structures
at $f^{GW}\sim f_{*1}^{S}$ and $\sim f_{*2}^{S}$ for relatively
large $r$. Such a principle peak-like configuration could possibly
have localized fine features with multiple small spikes (see Fig.\ref{fig-GW-RD}). The  RD GW spectrum at $\sim10^{-2}\textrm{Hz}$ region
can be probed by LISA/DECIGO/BBO for all cases here.
For $r=10$ case, it lies into the BBO sensitivity
limit at $\sim10^{-1}\textrm{Hz}$ region (upper-right panel), but it is far below the sensitivity
of next future GW detectors for $r=100,1000$ cases (lower panels).}

\label{fig-Cos-imp-GW1}
\end{figure}

\begin{figure}
\begin{centering}
\includegraphics[scale=0.96]{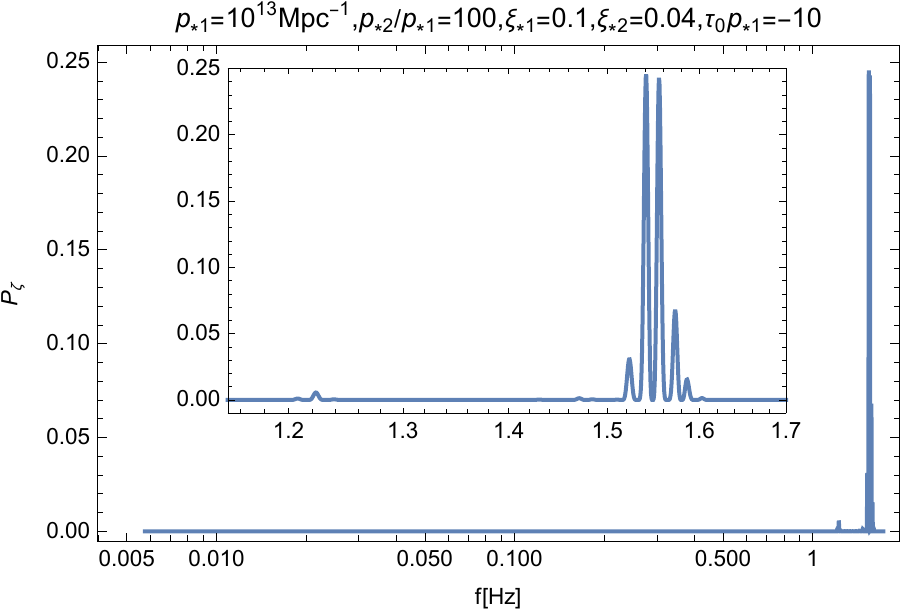}\includegraphics[scale=0.96]{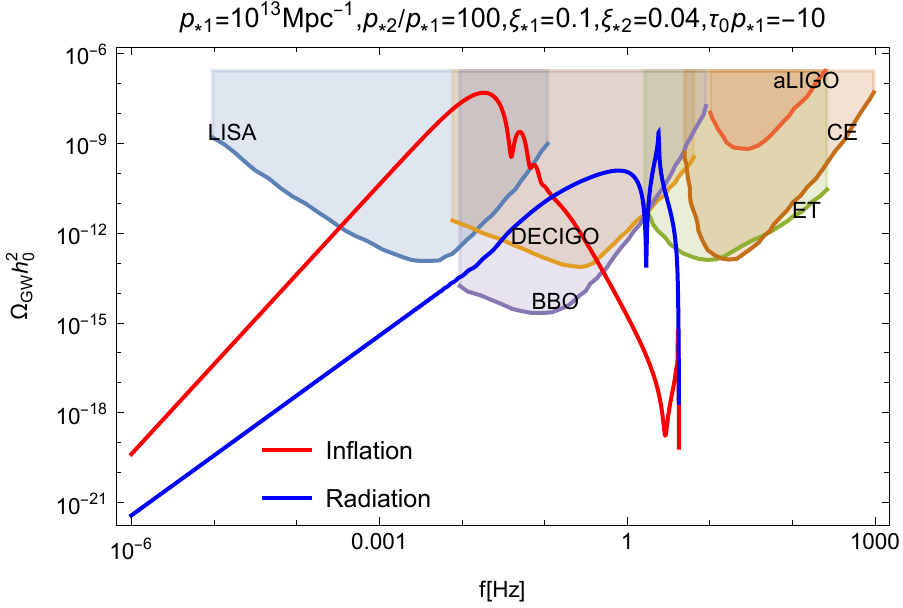}
\par\end{centering}
\begin{centering}
\includegraphics[scale=0.96]{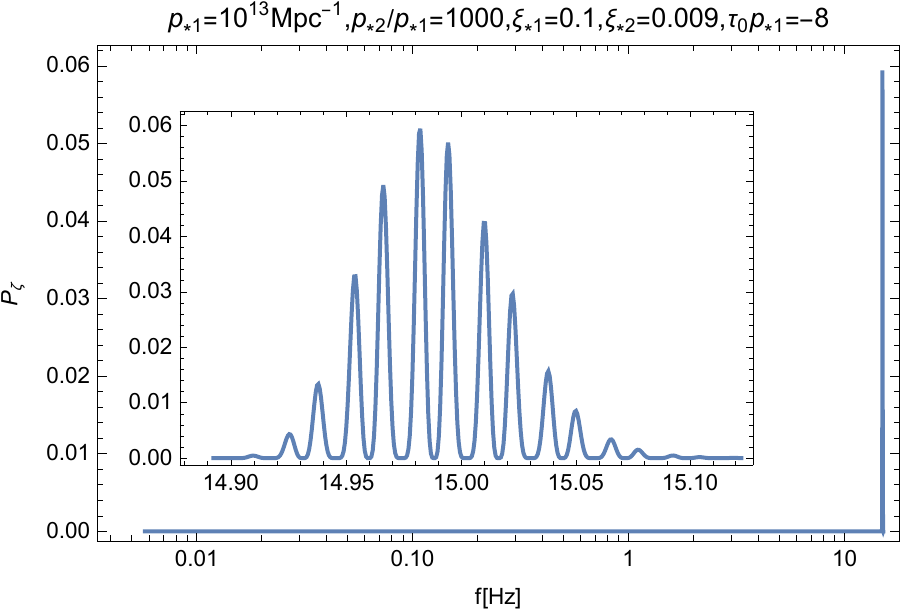}\includegraphics[scale=0.96]{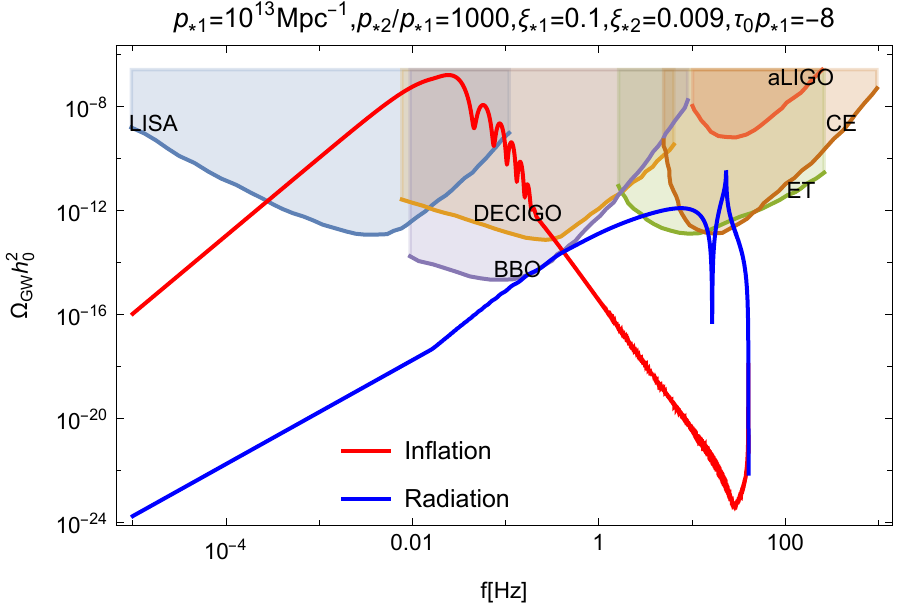}
\par\end{centering}
\caption{The power spectrum of the primordial curvature perturbations, for
relatively small value of dimensionless combination $p_{*1}\tau_{0}$ (left column),
and the corresponding GW energy spectrum compared with LISA, DECIGO, BBO,
aLIGO, ET and CE (right column). We take the fiducial scale $p_{*1}=10^{13}\textrm{Mpc}{}^{-1}$,
i.e. $f_{*1}^{S}\simeq10^{-2}\textrm{Hz}$, and display the cases
with $r=100$ (upper row) and $r=1000$ (lower row). Differently to 
 patterns shown in Fig.\ref{fig-ps} and Fig.\ref{fig-Cos-imp-GW1}, one can see
that the $f_{*1}^{S}$-peak group in the source spectrum as well as the
peak-like structure in RD GW spectrum at $f^{GW}\sim f_{*1}^{S}$ become invisible;
meanwhile the $\mathcal{O}(0.1)$ scale of dominant peaks in $f_{*2}^{S}$-peak
group significantly amplify the RD GW spectrum at $f^{GW}\sim f_{*2}^{S}$.
The resulted single peak RD GW spectrum now can be probed in the sensitivity region
of DECIGO/BBO (ET/CE) at $f^{GW}\sim f_{*2}^{S}\sim1\textrm{Hz}(10\textrm{Hz})$
for $r=100(1000)$. On the other hand, the major features of inflationary
GW spectrum almost remain the same 
with only a tiny shift of overall scale (compare to Fig.\ref{fig-Cos-imp-GW1}), namely the main broad peak is
now shifted to $f^{GW}\sim0.01\textrm{Hz}$, still visible for LISA. }

\label{fig-Cos-imp-GW2}
\end{figure}

\begin{figure}
	\begin{centering}
		\includegraphics[scale=0.96]{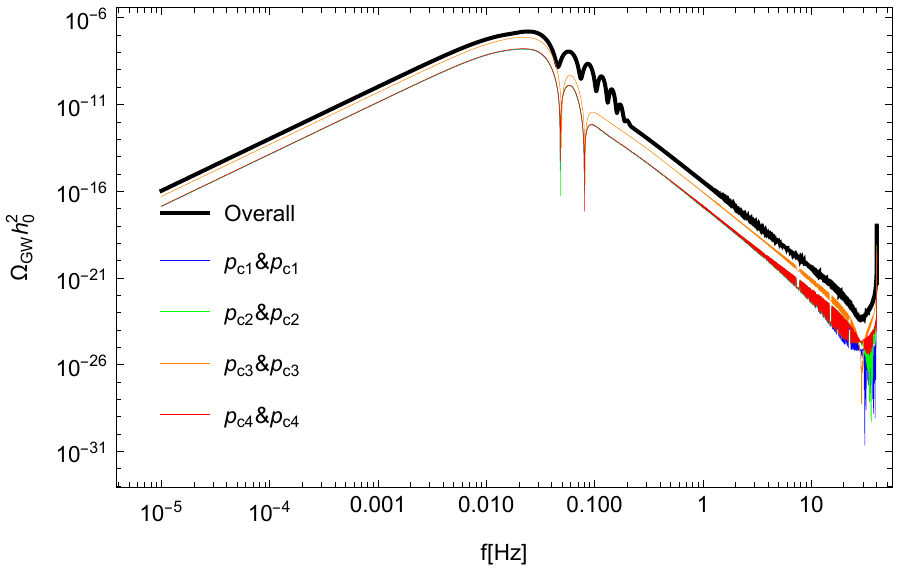}\includegraphics[scale=0.96]{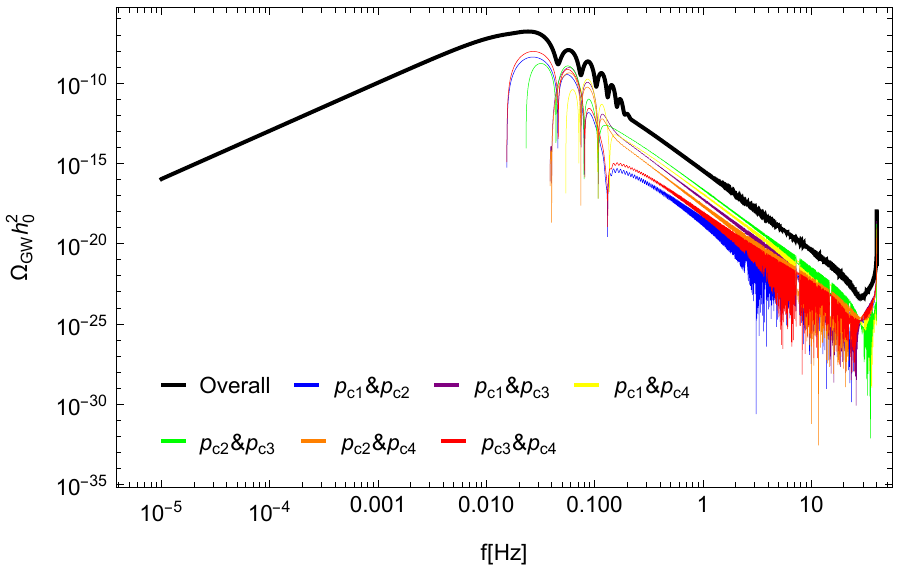}
		\par\end{centering}
	\caption{ The energy spectrum of induced GWs from inflationary era with parameter set as in lower panels in Fig. \ref{fig-Cos-imp-GW2}, namely $p_{*1}=10^{13}\textrm{Mpc}{}^{-1}, p_{*2}/p_{*1}=1000, \xi _{*1}=0.1, \xi_{*2}=0.009, \tau_0p_{*1}=-8$. The black line denotes the overall GW energy spectrum, and the other colored lines denote the decomposed components (see Eq. (\ref{eq:ph-inf})) generated by the convolution of several representative dominant resonating $p_{ci}$-modes displayed in the lower-left panel in  Fig. \ref{fig-Cos-imp-GW2}.  Similar to the analysis in Figs. \ref{fig-GW-Inf} and \ref{fig-En-Os}, both the parametric resonances of  envelope oscillatory modulation of each $p_{ci}$-mode (left panel) and intersection effect of the different $p_{ci}\& p_{cj}$-modes  (right panel) make contributions to the oscillatory pattern in the intermediate band of the  overall inflationary GWs spectrum. Moreover, the appearance of such regular GW oscillation (black line)  is also sensitive to the particular oscillatory profile of multiple dominant resonating peaks  in $P_{\zeta}$ shown in the lower-left panel in  Fig. \ref{fig-Cos-imp-GW2}.}
	
	\label{fig-Cos-imp-GW-osci-decomp}
\end{figure}

In this section, we will compare GW signals generated from inflationary and RD epochs with 
GW experiments \cite{Moore:2014lga}, e.g., the Laser Interferometer
Space Antenna (LISA) \cite{amaro2017laser}, DECi-hertz Interferometer
Gravitational Wave Observatory (DECIGO) \cite{kawamura2006japanese,Yagi:2011wg},
BigBang Observer (BBO) \cite{Corbin:2005ny,harry2006laser}, Advanced
LIGO + Virgo collaboration \cite{LIGOScientific:2016jlg,KAGRA:2013rdx},
Einstein Telescope (ET) \cite{punturo2010einstein,Sathyaprakash:2012jk}
and Cosmic Explorer (CE) \cite{LIGOScientific:2016wof,Reitze:2019iox}.
For a clear comparison of theory with the sensitivity curves of  GW observations,
we use the GW frequency $f^{GW}/\textrm{Hz}$ instead of the wave-number
$k/\textrm{Mpc}{}^{-1}$ by the relation $f^{GW}=1.55\times10^{-15}\left(\frac{k}{1\textrm{Mpc}{}^{-1}}\right)\textrm{Hz}$.
For convenience,  we introduce the frequency of the scalar source mode
$f^{S}/\textrm{Hz}$ related to its wave-number $p/\textrm{Mpc}{}^{-1}$
by $f^{S}=1.55\times10^{-15}\left(\frac{p}{1\textrm{Mpc}{}^{-1}}\right)\textrm{Hz}$.
Accordingly, we provide a dictionary among the characteristic scales
$p_{*1,2}$ and $f_{*1,2}^{S}$, the dominant resonating modes $p_{ci}$
and $f_{ci}^{S}$, and so on. Let us notice that all previous results above
can be easily regained in terms of $f^{GW,S}$. On the other hand,
due to the scaling invariance of the GW energy spectrum mentioned
below Eq.\eqref{eq:es-inf}, we choose to fix one of the characteristic
wave-number $p_{*1}$ in double cosine $c_s$ parameterization to be $p_{*1}=10^{13}\textrm{Mpc}{}^{-1}$,
i.e. $f_{*1}^{S}\simeq10^{-2}\textrm{Hz}$, which falls in the optimal
sensitivity window of LISA. Employing Eqs.\eqref{eq:GW-RD}, \eqref{eq:ph-inf}
and \eqref{eq:es-inf} for different ratios $r\equiv f_{*2}^{S}/f_{*1}^{S}$,
we show a comparison among the induced GW abundance from inflationary and RD phases 
with the sensitivity curves of various GW observations in Figs.\ref{fig-Cos-imp-GW1}
and \ref{fig-Cos-imp-GW2}. We mention that in all cases, the conditions are 
$P_{\zeta}<1$ and $P_{h}<1$, or equivalently $\Omega_{\mathrm{GW}}(\tau_{0},f)h_{0}^{2}\lesssim10^{-6}$,
avoiding the loss of the perturbative control for both
the inflationary and RD eras.

In Fig.\ref{fig-Cos-imp-GW1}, the parameters in the first three
panels are simply rescaled of those in Fig.\ref{fig-ps},
indicating that the profile of the corresponding source spectrum $P_{\zeta}$ remains
the same except a overall shift in the scale of $p$ (or $f^{S}$) due to
scale invariance. More precisely, in the $(P_{\zeta}, f^{S})$ plane
after rescaling, there exists one dominant resonating peak in the ``$f_{*1}^{S}$-peak
group'' (similar definition as ``$p_{*1}$-peak group'' given
in Sec. \ref{sec:Sound-Speed-Resonance}) centered at $\sim f_{*1}^{S}$,
and usually multiple dominant peaks in the ``$f_{*2}^{S}$-peak group''
centered at $\sim f_{*2}^{S}$ . In Fig.\ref{fig-Cos-imp-GW1}, one
can see that the magnitude of the present GW energy spectrum from
inflation era can be comparable to or even larger than that from the
RD era. As discussed in Sec. \ref{sec:Induced-Gravitational-Waves},
the major broad peak of the GW spectrum from inflation is located
at $f^{GW}\sim f_{ci}^{S}\xi_{i}\sim f_{*1}^{S}\xi_{*1}\sim f_{*2}^{S}\xi_{*2}\sim10^{-3}\textrm{Hz}$
in all cases shown in Fig.\ref{fig-Cos-imp-GW1}, which probes the
sensitivity range of LISA. In the tail of the inflationary GW spectrum,
multiple spikes fall within the sensitivity band of DECIGO/BBO in
the $\left(10^{-2}\textrm{Hz},10^{-1}\textrm{Hz}\right)$ region,
generated by the oscillatory modulation of the envelope
(see Fig.\ref{fig-En-Os}) and the convolution of different resonating
source modes (see Fig.\ref{fig-GW-Inf}). However, for the induced
GW spectrum from RD era, its primary profile is sensitive to the
positions and amplitudes of the dominant multiple peaks in source spectrum (see
Figs.\ref{fig-ps} and \ref{fig-GW-RD}). In Fig.\ref{fig-Cos-imp-GW1},
GW spectrum exhibits only a principle peak-like structure at $f^{GW}\sim f_{*1}^{S}\sim10^{-2}\textrm{Hz}$
for small ratio $r=2$, whereas two principle peak-like structures
at $f^{GW}\sim f_{*1}^{S}\sim10^{-2}\textrm{Hz}$ and $f^{GW}\sim f_{*2}^{S}\sim r\times10^{-2}\textrm{Hz}$
for relatively large ratios $r=10,100,1000$, respectively. On top
of the principal peak-like structure, localized features with multiple
small spikes can appear as shown in Fig.\ref{fig-GW-RD}.
We find that the RD GW spectrum at $f^{GW}\sim f_{*1}^{S}\sim10^{-2}\textrm{Hz}$
band lies in the sensitivity regime of LISA/DECIGO/BBO for all cases in
Fig.\ref{fig-Cos-imp-GW1} and at $f^{GW}\sim f_{*2}^{S}\sim r\times10^{-2}\textrm{Hz}$ band
the BBO sensitivity limit for $r=10$ (upper-right panel); while
far below the signal sensitivity available to the next generation
of detector for $r=100,1000$ (lower panels). In fact, to ensure the
detectability of the RD GW peaks at LISA frequency window, the dimensionless
combination $p_{*1}\tau_{0}$ controlling the height of the dominant
peak in the $f_{*1}^{S}$-peak group is set to be large enough for
 cases in Fig.\ref{fig-Cos-imp-GW1}. Consequently, $\xi_{*2}$
shall be small enough to meet the constraints $P_{\zeta}<1$
and $P_{h}<1$, which render the $f_{*2}^{S}$-peak group in source
spectrum (the principle peak-like structure at $f^{GW}\sim f_{*2}^{S}$
in RD GW spectrum) lower than the $f_{*1}^{S}$-peak group (the
principle peak-like structure at $f^{GW}\sim f_{*1}^{S}$ in RD GW spectrum) of around $2\sim4$
($4\sim16$) orders of magnitude for large $r$. Therefore, in order to increase
the magnitude of RD GW band at $\sim f_{*2}^{S}$ for $r=100,1000$,
the price to pay is to reduce the amplitude of the $f_{*1}^{S}$-peak
group of source spectrum, and hence decreasing the RD GW abundance at $f^{GW}\sim f_{*1}^{S}$.
Indeed, as displayed in Fig. \ref{fig-Cos-imp-GW2}, by taking much
smaller $p_{*1}\tau_{0}$, the $f_{*1}^{S}$-peak group in source
spectrum as well as the corresponding peak in RD GW spectrum become
invisible, meanwhile the dominant peaks in $f_{*2}^{S}$-peak group
increase to the $\mathcal{O}(0.1)$ scale, leading to a significant
amplification of RD GW abundance at $f^{GW}\sim f_{*2}^{S}$ band.
For $r=100$ the RD GW spectrum peak  at $f^{GW}\sim f_{*2}^{S}\sim1\textrm{Hz}$
can be testable in DECIGO/BBO, and for $r=1000$ it is expected to be visible in
the ET/CE band at $f^{GW}\sim10\textrm{Hz}$. On the
other hand, comparing  Fig. \ref{fig-Cos-imp-GW1} and Fig. \ref{fig-Cos-imp-GW2}, we find that the main features of
inflationary GW spectrum are almost unchanged with only a slight
overall shift and the major broad peak (now at $\sim0.01\textrm{Hz}$)
is still visible for LISA. Let us notice that here
the multiple spikes in the intermediate band of the inflationary GW spectrum have same origins as those in Fig. \ref{fig-Cos-imp-GW1}, namely from the intersection effect of the multiple resonating peaks and the parametric resonances of the envelope oscillatory modulation(see Fig. \ref{fig-Cos-imp-GW-osci-decomp}).

Furthermore, thanks to the scale invariance, both the overall GW spectra
from RD and inflationary epochs can be shifted along the frequency axis
by freely setting a fiducial frequency to $f_{*1}^{S}$. This implies
that the induced GWs can overlap with each sensitivity range of
all GW detectors. For instance, taking $p_{*1}=10^{7}\textrm{Mpc}{}^{-1}$,
i.e. $f_{*1}^{S}\simeq10^{-8}\textrm{Hz}$, some certain major peaks
of RD and inflationary GW spectra can be at lower frequencis
around $f^{GW}\sim10^{-8}\textrm{Hz}$ in 
the sensitive window accessible by the ongoing and planned pulsar
timing array experiments such as the North American Nanohertz Observatory
for Gravitational Waves (NANOGrav) \cite{NANOGrav:2020bcs,DeLuca:2020agl},
Square Kilometre Array (SKA) \cite{Janssen:2014dka} and International
Pulsar Timing Array (IPTA) \cite{hobbs2010international}.

\section{Discussion and Conclusions}

\label{sec:Discussion-and-conclusion}

In this paper, we investigated the parametric resonance effects from the
modified time-varying sound speed with the double oscillatory behavior. 
Sound speed resonances 
enhance the primordial density perturbations and hence amplify the abundance of the second order GWs induced during inflationary and RD epochs. 
We performed a comprehensive
analysis from both theoretical and phenomenological sides. The resonant modes manifest themselves 
as narrow and sharp peaks in the power spectrum of the primordial
density perturbations, amplifying the production of  GWs during the inflationary and RD eras.  
GWs from early Universe dynamics survive as a characteristic 
 stochastic background at the present day. 
 We showed that a {\it resonance gravitational memory} of sound-speed oscillations during inflation is carried by the relic stochastic GW background and it
can be detected by next generation of GW interferometers. 
 The
presence of two oscillatory modes in the sound speed yields 
interesting results. The two characteristic frequencies
in Eq.\eqref{eq:double-cs} generically lead to multiple resonating
peaks distributed in two groups centered around $p_{*1}$ and $p_{*2}$ in $P_{\zeta}$ (see
Figs.\ref{fig-vsol} and \ref{fig-ps}), except some extreme cases
like vanishing $\beta_{*2}$ (see the upper-left panel in Fig.\ref{fig-ps})
and very small value of $p_{*1}\tau_{0}$ (see the left column in
Fig.\ref{fig-Cos-imp-GW2}). The appearance of multiple peaks, especially
the dominant peak(s) in each group, characterize 
the spectrum of the induced GWs. Specifically, for the parameter space
within the perturbative regime, there exists a single broad peak of
the GW spectrum from inflation, the tails of which consist of multiple
secondary spikes. The secondaries are generated by the convolution of multiple dominant
peaks in source spectrum as well as the oscillatory modulation of certain resonating
modes' envelope (see Figs.\ref{fig-GW-Inf}-\ref{fig-En-Os}).
On the other hand, GW spectrum from RD epoch exhibits one (two) principle
peak-like configuration at $\sim p_{*1}$ ($\sim p_{*1}$ and $\sim p_{*2}$)
for relatively small (large) ratio $r=p_{*2}/p_{*1}$, which, in certain parametric regions, can
contain a fine structure of spikes from multiple peaks in source spectrum
(see Fig.\ref{fig-GW-RD}). Finally, we studied phenomenological
implications by fixing a particularly interesting benchmark frequency $p_{*1}=10^{13}\textrm{Mpc}{}^{-1}$.
We found that the major peak of the  inflationary GW spectrum
can be detected in LISA, meanwhile the principle peak-like structure of
RD GW spectrum at $\sim10^{-2}\textrm{Hz}$ overlaps with the sensitivity
window of LISA/DECIGO/BBO. The other principle peak-like
structure at $\sim r\times10^{-2}\textrm{Hz}$ results as sub-dominated by
at least $5$ orders of magnitude (see Fig.\ref{fig-Cos-imp-GW1}).
Moreover, in the extreme cases with very small $p_{*1}\tau_{0}$,
the abundance of RD GWs  at $\sim r\times10^{-2}\textrm{Hz}$ band
is largely boosted and can be detected in DECIGO/BBO or ET/CE (see
Fig.\ref{fig-Cos-imp-GW2}). In short, the distinctive behavior of
the present relic GW spectrum generated during inflationary and
RD epochs can be analyzed as 
complementary features of signals at different frequency windows.
Such a phenomena can be promisingly detected  in several current and next future 
GW experiments.

As it is well known, PBHs can be formed
when the primordial density perturbations of certain modes are dramatically
amplified after re-entering into the Hubble horizon \cite{zel1967hypothesis,hawking1971gravitationally}.
Accordingly, the presence of multiple resonating peaks in the spectrum
of curvature perturbations definitely increases the abundance of PBHs.
The amplitudes and positions of these peaks determine
the mass spectrum and fraction of PBHs, in turn  constrained
by several observations \cite{Carr:2009jm,Sasaki:2018dmp,Carr:2020gox}.
Therefore, searches for induced GWs and PBHs produced
in our model from electromagnetic
surveys and GW interferometers appear as promising in the era of multi-messenger
astronomy. On the other hand, the overly enhancement of the primordial
density perturbations at certain scale unavoidably induces non-Gaussianities
of relic GWs at non-linear order \cite{Thrane:2013kb,Cai:2018dig,Cai:2019amo,Adshead:2021hnm}.
Thus, multiple resonating primordial source modes
predicted in our model could imprint characteristic non-Gaussianities in the induced GW bispectrum.
We think that such an effect deserves future investigations beyond the purposes of this work. 

We also mentioned above that the double cosine parameterization
of the sound speed could be extended to a more general scenario as Eq. \eqref{eq:n-cos-cs} 
with $N$ characteristic oscillation modes. From the resonance patterns of the single and double cosine scenarios, we can envisage that,  generically,
there will be up to $N$ groups of resonating peaks centered at $\sim p_{*1},p_{*2},\cdots,p_{*N}$ in the power spectrum of primordial curvature perturbations. 
For the parameter space within perturbative regime, one plausible result is that  a single major broad peak appears in the inflationary GW spectrum, whereas $N$ principle peak-like structures in the RD GW spectrum. In addition, in Refs.\cite{Palma:2020ejf,Fumagalli:2020adf,Fumagalli:2020nvq}
the authors showed that the primordial scalar power spectrum at small
scales can be exponentially enhanced in the multi-field inflationary
paradigm with a sharp pivot in the inflationary trajectory.
Usually, inflaton field(s) undergoes to strong oscillations after a
 sudden turn aforementioned. Thus it would be an intriguing possibility to study a related sound speed
resonance mechanism in multifield inflationary scenarios \cite{Chen:2011zf,Gao:2013ota}.
Furthermore, the power spectrum of primordial curvature perturbations
exhibits oscillations in several frequency intervals, for certain parameter choices (e.g., the insets in Fig.\ref{fig-Cos-imp-GW2}).
Such cases can be considered as related to the spectra with particular sharp
features as ones displayed in Fig.9 in Ref.\cite{Fumagalli:2020nvq}.
Indeed, the spectrum of  GWs sourced during inflationary
and RD era would inherit richer localized oscillatory features
superposing on the principal peak-like structures, beyond the delta-peak approximation adopted in this paper (see
Eqs.\eqref{eq:fit-source-ps}, \eqref{eq:GW-RD} and \eqref{eq:ph-inf}).
Finally, in this work we have observed an interesting phenomena:
the sinusoidal modulation of the resonating modes' envelope boosts
the energy density of GWs at a narrow band (see Fig.\ref{fig-En-Os}).
This could be considered as another kind of parametric resonance effect that 
deserves future investigations.

\vspace{2cm}

\noindent \textbf{Acknowledgements.}
We thank Peng Wang, Shengfeng Yan, Chao Chen and Guangzhou Guo for their helpful discussions and suggestions.  Q.Y. Gan work is supported by the scholarship from China Scholarship Council (CSC) under the Grant CSC No. 202106240085. 
A.A. work is supported by the Talent Scientific Research Program of College of Physics, Sichuan University, Grant No.1082204112427 \& the Fostering Program in Disciplines Possessing Novel Features for Natural Science of Sichuan University,  Grant No. 2020SCUNL209 \& 1000 Talent program of Sichuan province 2021. 
S.C. acknowledges the support of Istituto Nazionale di Fisica Nucleare sez. di Napoli {\it (iniziative specifiche} QGSKY and Moonlight2).

\bibliographystyle{unsrturl}

\end{document}